\documentclass[reprint,amsmath, amssymb, aps, prl, citeautoscript, superscriptaddress, nobibnotes, longbibliography]{revtex4-2}

\usepackage{soul}
\usepackage{siunitx}
\usepackage{graphicx}
\usepackage{braket}
\usepackage[colorlinks]{hyperref}
\usepackage[capitalize]{cleveref}
\usepackage[normalem]{ulem}
\usepackage{comment}

\hypersetup{citecolor={black}}  
\setcitestyle{super}
\usepackage{float}

\newcommand{\nodea}{node A}

\newcommand{\eup}{$\ket{\uparrow_{\text{e}}}$}
\newcommand{\edown}{$\ket{\downarrow_{\text{e}}}$}
\newcommand{\eupA}{$\ket{\uparrow^{A}_{\text{e}}}$}
\newcommand{\edownA}{$\ket{\downarrow^{A}_{\text{e}}}$}

\newcommand{\nup}{$\ket{\uparrow_{\text{n}}}$}
\newcommand{\ndown}{$\ket{\downarrow_{\text{n}}}$}
\newcommand{\ebin}{$\ket{e}$}
\newcommand{\lbin}{$\ket{l}$}
\newcommand*{\citen}[1]{%
  \begingroup
    \romannumeral-`\x %
    \setcitestyle{numbers}%
    \cite{#1}%
  \endgroup   
}

\makeatletter
\newcommand*{\balancecolsandclearpage}{%
  \close@column@grid
  \newpage
  \twocolumngrid
}
\makeatother

\makeatletter
\newcommand*{\balancecolsandclearpagesingle}{%
  \close@column@grid
  \newpage
}
\makeatother

\newcounter{EDfig}
\crefname{EDfig}{Extended Data Fig.}{Extended Data Fig.}

\newcounter{SIfig}
\crefname{SIfig}{Fig.}{Fig.}
\Crefname{SIfig}{Fig.}{Fig.}

\newcounter{SItable}
\crefname{SItable}{Table}{Table}
\Crefname{SItable}{Table}{Table}

\begin{document}

\title{Entanglement of Nanophotonic Quantum Memory Nodes \\ in a Telecom Network}

\author{C. M. Knaut}
\thanks{These authors contributed equally to this work.}
\affiliation{Department of Physics, Harvard University, Cambridge, Massachusetts 02138, USA}
\author{A. Suleymanzade}
\thanks{These authors contributed equally to this work.}
\affiliation{Department of Physics, Harvard University, Cambridge, Massachusetts 02138, USA}
\author{Y.-C. Wei}
\thanks{These authors contributed equally to this work.}
\affiliation{Department of Physics, Harvard University, Cambridge, Massachusetts 02138, USA}
\author{D. R. Assumpcao}
\thanks{These authors contributed equally to this work.}
\affiliation{John A. Paulson School of Engineering and Applied Sciences, Harvard University, Cambridge, Massachusetts 02138, USA}
\author{P. -J. Stas}
\thanks{These authors contributed equally to this work.}
\affiliation{Department of Physics, Harvard University, Cambridge, Massachusetts 02138, USA}
\author{Y. Q. Huan}
\affiliation{Department of Physics, Harvard University, Cambridge, Massachusetts 02138, USA}
\author{B. Machielse}
\affiliation{Department of Physics, Harvard University, Cambridge, Massachusetts 02138, USA}
\affiliation{AWS Center for Quantum Networking, Boston, Massachusetts 02210, USA}
\author{E. N. Knall}
\affiliation{John A. Paulson School of Engineering and Applied Sciences, Harvard University, Cambridge, Massachusetts 02138, USA}
\author{M. Sutula}
\affiliation{Department of Physics, Harvard University, Cambridge, Massachusetts 02138, USA}
\author{G. Baranes}
\affiliation{Department of Physics and Research Laboratory of Electronics, Massachusetts Institute of Technology, Cambridge, MA 02139, USA}
\affiliation{Department of Physics, Harvard University, Cambridge, Massachusetts 02138, USA}
\author{N. Sinclair}
\affiliation{John A. Paulson School of Engineering and Applied Sciences, Harvard University, Cambridge, Massachusetts 02138, USA}
\author{C. De-Eknamkul}
\affiliation{AWS Center for Quantum Networking, Boston, Massachusetts 02210, USA}
\author{D. S. Levonian}
\affiliation{Department of Physics, Harvard University, Cambridge, Massachusetts 02138, USA}
\affiliation{AWS Center for Quantum Networking, Boston, Massachusetts 02210, USA}
\author{M. K. Bhaskar}
\affiliation{Department of Physics, Harvard University, Cambridge, Massachusetts 02138, USA}
\affiliation{AWS Center for Quantum Networking, Boston, Massachusetts 02210, USA}
\author{H. Park}
\affiliation{Department of Physics, Harvard University, Cambridge, Massachusetts 02138, USA}
\affiliation{Department of Chemistry and Chemical Biology, Harvard University, Cambridge, Massachusetts 02138, USA}
\author{M. Lon\v{c}ar}
\affiliation{John A. Paulson School of Engineering and Applied Sciences, Harvard University, Cambridge, Massachusetts 02138, USA}
\author{M. D. Lukin}
\email{lukin@physics.harvard.edu }
\affiliation{Department of Physics, Harvard University, Cambridge, Massachusetts 02138, USA}

\date{May 15, 2024}

\begin{abstract}
A key challenge in realizing practical quantum networks for long-distance quantum communication involves robust entanglement between quantum memory nodes connected via fiber optical infrastructure \cite{Kimble2008, Briegel1998, Childress_2006}. Here, we demonstrate a two-node quantum network composed of multi-qubit registers based on silicon-vacancy (SiV) centers in nanophotonic diamond cavities integrated with a telecommunication (telecom) fiber network. Remote entanglement is generated via the cavity-enhanced interactions between the SiV's electron spin qubits and optical photons. Serial, heralded spin-photon entangling gate operations with time-bin qubits are used for robust entanglement of separated nodes.  Long-lived nuclear spin qubits are used to provide second-long entanglement storage and integrated error detection. By integrating efficient bi-directional quantum frequency conversion of photonic communication qubits to telecom frequencies (1350 nm), we demonstrate entanglement of two nuclear spin memories through 40 km spools of low-loss fiber and a 35 km long fiber loop deployed in the Boston area urban environment, representing an enabling step towards practical quantum repeaters and large-scale quantum networks.
\end{abstract}

\maketitle

\begin{figure*}[htp]
    \centering
     \includegraphics[width=1\linewidth]{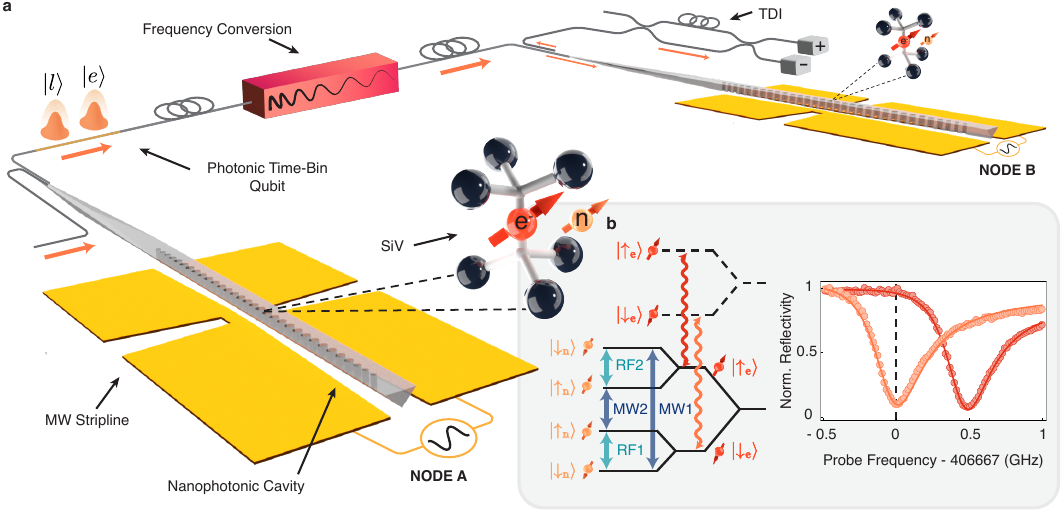} 
    \caption{\textbf{A two-node quantum network of cavity-coupled solid state emitters.} \textbf{a.} Experimental setup. Each SiV is localized in a nanophotonic cavity within an individually operated cryostat held at temperatures below \SI{200}{\milli \kelvin} in two separate laboratories. The line-of-sight distance between the two SiVs is \SI{6}{\meter}.  A gold coplanar waveguide is used to deliver MW and RF pulses to the SiV. Both quantum network nodes are connected via a $\approx \SI{20}{\meter}$ long optical fiber and frequency shifting setup to compensate for differences in the optical transition frequencies, or a long telecom fiber link using QFC (see \cref{fig:4} a). The measurement of the photonic time-bin qubit is performed at node B using a time-delay interferometer (TDI), which measures the time-bin qubit in the basis $\ket{\pm} \propto (\ket{e} \pm \ket{l})$. \textbf{b.} Left: Energy levels of $^{29}$SiV showing the MW and RF transitions in the two-qubit manifold (blue and turquoise arrows) and the spin-conserving optical transitions (red and orange). Right: The reflection spectrum of node A's cavity QED system shows the electron-spin-dependent cavity reflectance. The dashed line indicates the frequency of maximum reflectance contrast, which is used as the frequency for the electron spin state readout and the photonic entanglement.}
\label{fig:1}
\end{figure*}

Distributing quantum  entanglement between quantum memory nodes separated by 
extended distances \cite{Kimble2008, doi:10.1126/science.aam9288} is a critical element for the realization of quantum networks, enabling potential applications ranging from quantum repeaters  \cite{Briegel1998, azuma2023quantum} and long-distance secure communication \cite{RevModPhys.74.145, PhysRevLett.67.661} to distributed quantum computing \cite{Monroe_2014, ang2022architectures} and distributed quantum sensing and metrology \cite{Khabiboulline_2019PRL, Giovannetti_2001}.  
Proposed architectures require quantum nodes containing multiple long-lived qubits that can collect, store, and process information communicated via photonic channels based on telecommunication (telecom) fibers or satellite-based links. In particular, the abilities to herald on successful photon arrival events and to detect quantum-gate errors are central for scalable implementations. Since photons and individual matter qubits interact weakly in free space \cite{RevModPhys.87.1379},  a promising approach to enhance the interaction between light and communication qubits is to utilize nanophotonic cavity quantum electrodynamic (QED) systems, where tight light confinement inside the nanostructure enables strong interactions between the photon and the communication qubit \cite{doi:10.1126/science.aah6875, Nguyen2019, ourari2023indistinguishable, Ruskuc_2022}. Furthermore, nanophotonic systems offer a path towards large-scale manufacturing and on-chip electric and optical control integration \cite{riedel2023efficient, Molesky2018, parker2023diamond}. Several experiments demonstrated remote entanglement in systems ranging from neutral atoms \cite{hofmann2012heralded, Daiss_2021, Ritter2012AnEQ, van_Leent_2022} and trapped ions \cite{PhysRevLett.130.050803, PhysRevLett.124.110501} to semiconductor quantum dots \cite{PhysRevLett.119.010503} and nitrogen-vacancy (NV) centers in diamond \cite{Bernien_2013, Humphreys_2018}.  Recently, two atomic ensemble memories have been entangled through a metropolitan fiber network \cite{PhysRevLett.129.050503, Yu_2020, liu2023multinode}. However, real-world applications require a combination of efficient photon coupling, long-lived heralded memory, and multi-qubit operations with practical telecom fiber networks, which is an outstanding challenge. \\

Here, we report the realization of a two-node quantum network between two multi-qubit quantum network nodes constituted by silicon-vacancy (SiV) centers in diamond coupled to nanophotonic cavities and integrated with a telecom fiber network. SiVs coupled to cavities have emerged as a promising quantum network platform, having demonstrated memory-enhanced quantum communication \cite{Bhaskar2020} and robust multi-qubit single-node operation \cite{doi:10.1126/science.add9771}. We extend these single-node experiments by demonstrating remote entanglement generation between two electron spins in two spatially separated SiV centers with a success rate up to \SI{1}{\hertz}. Our approach utilizes serial, heralded spin-photon gate operations with time-bin qubits for robust entanglement of separated nodes and does not require phase stability across the link. We further make use of the multi-qubit capabilities to entangle two long-lived nuclear spins, using integrated error detection to enhance entanglement fidelities and dynamical decoupling sequences to extend the entanglement duration to \SI{1}{\second}. Both entanglement generation techniques rely on the strong light-matter interaction enabled by the SiV's coupling to the nanophotonic cavity.  In order to demonstrate the feasibility of deployed quantum networks using our platform, we use bi-directional quantum frequency conversion (QFC) to convert the photonic qubits' wavelength to telecom wavelengths. Building on recently demonstrated compatibility of our platform with bi-directional QFC \cite{bersin2023telecom, bersin2023development}, we demonstrate remote entanglement generation through spools of up to \SI{40}{\kilo \meter} of low-loss telecom fiber. Finally, we combine these techniques to demonstrate entanglement generation through a \SI{35}{\kilo \meter}-long loop of fiber with 17 dB loss deployed in the Boston area urban environment.

\subsection{Two-node quantum network using integrated nanophotonic systems}

Our quantum network nodes consist of SiV centers in diamond that reside in individually operated dilution refrigerator setups in separate laboratories (\cref{fig:1} a). By selectively implanting the $^{29}$Si isotope into the diamond substrate,  each SiV deterministically contains two addressable spin qubits: one electron spin used as a communication qubit, which couples strongly to itinerant photons, and one long-lived $^{29}$Si nuclear spin, used as a memory qubit to store entanglement. Under an externally applied magnetic field, Zeeman sublevels define the electronic spin qubits states (\edown{}, \eup{}) and the nuclear spin qubit states (\ndown{}, \nup{})\cite{Hepp2014a, Pingault2017} (\cref{fig:1} b, left). Microwave (MW) pulses are used to drive the electronic spin-flipping transitions, while radio-frequency (RF) pulses drive the nuclear spin-flipping transitions \cite{doi:10.1126/science.add9771}. The SiV centers are embedded into nanophotonic diamond cavities, which enhance interactions between light and the electron spin \cite{RevModPhys.87.1379, Duan2004}. The strong emitter-cavity coupling as characterized by the single-photon cooperativity in node A (node B) of 12.4 (1.5) \cite{SI} results in an electron-spin dependent cavity reflectance (\cref{fig:1} b, right) \cite{Nguyen2019}.   This can be utilized to construct a reflection-based spin-photon gate ($e-\gamma$ gate), which contains a sequence of rapid MW gates generating entanglement between the SiV's electron spin and the photonic qubits \cite{Nguyen2019}. Additionally, taking advantage of the strong coupling between the SiV's electron spin and the $^{29}$Si nuclear spin, nucleus-photon entanglement can be created using the PHOton-Nucleus Entangling (PHONE) gate as demonstrated recently \cite{ doi:10.1126/science.add9771}. The two nodes are either connected directly via a $\approx \SI{20}{\meter}$ long optical fiber (\cref{fig:1} a), or via a considerably longer telecom fiber link as discussed below (\cref{fig:4} a).

We use a serial network configuration to generate remote entanglement between the electron spins in node A and node B, mediated by a time-bin photonic qubit (\cref{fig:2} a). We first use a $e-\gamma$ gate to generate an entangled Bell state between \nodea{}'s electron spin \edownA{}, \eupA{} and an incoming time-bin photonic qubit \ebin{}, \lbin{}  \cite{Nguyen2019}. Here, \ebin{} and \lbin{} describe the presence of a photon in the early and late time-bins of the photonic qubit, which are separated by $\delta t = \SI{142}{\nano \second}$, respectively. The resulting photon-electron Bell state can be described as $\ket{\text{Photon}, \text{SiV A}} = (\ket{e \downarrow^{A}_{\text{e}}} + \ket{l \uparrow^{A}_{\text{e}}}) / \sqrt{2}$ \cite{Methods}.
After that, the photonic qubit travels via optical fiber to node B, where a second $e-\gamma$ gate entangles the photonic qubit with the electron spin in node B. In the ideal, lossless case, the resulting state is a three-particle  Greenberger–Horne–Zeilinger (GHZ) state:

\begin{align*}
\ket{\text{Photon}, \text{SiV A}, \text{SiV B}} &= (\ket{e \downarrow^{A}_{\text{e}}  \downarrow^{B}_{\text{e}}} + \ket{l \uparrow^{A}_{\text{e}} \uparrow^{B}_{\text{e}}}) / \sqrt{2}\\
&=  \left(\ket{+} \ket{\Phi_{\text{ee}}^+} + \ket{-} \ket{\Phi_{\text{ee}}  ^-}\right) / \sqrt{2}.
\end{align*}

Here, $\ket{\pm} = (\ket{e} \pm \ket{l})/ \sqrt{2}$ describes two orthogonal superposition states of the photonic time-bin qubit, and $\ket{\Phi_{\text{ee}}^\pm} = (\ket{ \downarrow^{A}_{\text{e}}  \downarrow^{B}_{\text{e}}} \pm \ket{ \uparrow^{A}_{\text{e}} \uparrow^{B}_{\text{e}}}) / \sqrt{2}$ describes the maximally entangled Bell states of the two spatially separated electron spins. The photonic qubit is measured in the $\ket{\pm}$-basis using a TDI to herald the generation of an electronic Bell state:

\begin{equation*}
\ket{\text{SiV A}, \text{SiV B}} = 
\begin{cases}
    \ket{\Phi_{\text{ee}}^+} ,& \text{if TDI measures} \ket{+} \\
    \ket{\Phi_{\text{ee}}^-} ,& \text{if TDI measures} \ket{-}.
\end{cases}
\end{equation*}

Note that similar to the previously utilized single-node schemes\cite{Nguyen2019}, this method is robust to photon loss, as any losses of photons can be detected by a missing heralding event. Furthermore, the main advantage of our serial scheme is that both the early and late time-bins of the photonic qubit travel through the same path, so no phase or polarization-locking is necessary to guarantee high interference contrast at the TDI. This relaxes the requirements on system stability compared to one-photon schemes, which typically require an interferometric measurement of two emitted photons traveling through two stabilized paths\cite{Humphreys_2018, liu2023multinode, PhysRevLett.119.010503, van_Leent_2022}, and avoids the reduction in entanglement rate typically present in two-photon schemes\cite{Barrett_2005, Bernien_2013}. Furthermore, extending the number of network nodes to more than two can be achieved by either connecting more than two nodes in series, or by employing a switch network between multiple nodes to generate pairwise connectivity. 
 
Since cavity-coupled $^{29}$SiV centers possess an inhomogeneous distribution of optical transition frequencies of around \SI{\pm 50}{\giga \hertz} centered around \SI{406.640}{\tera \hertz} (\SI{737.2}{\nano \meter}) \cite{Methods, srujanstrain}, the frequency difference between the nodes needs to be coherently bridged. For node B used in this work, for instance, the SiV's optical frequency $\omega_{B}$ is detuned from that of node A ($\omega_{A}$) by $\Delta_{\omega} = \SI{13}{\giga \hertz}$. To address this, we prepare the photonic qubit at frequency $\omega_{A}$ and then coherently shift its frequency by $\Delta_{\omega}$ after it has interacted with the SiV at node A, either using electro-optic frequency shifting or via bi-directional QFC \cite{bersin2023telecom, bersin2023development}. 

\subsection{Electronic spin entanglement}

\begin{figure}[hbt!]
    \centering
    \includegraphics[width=1\linewidth]{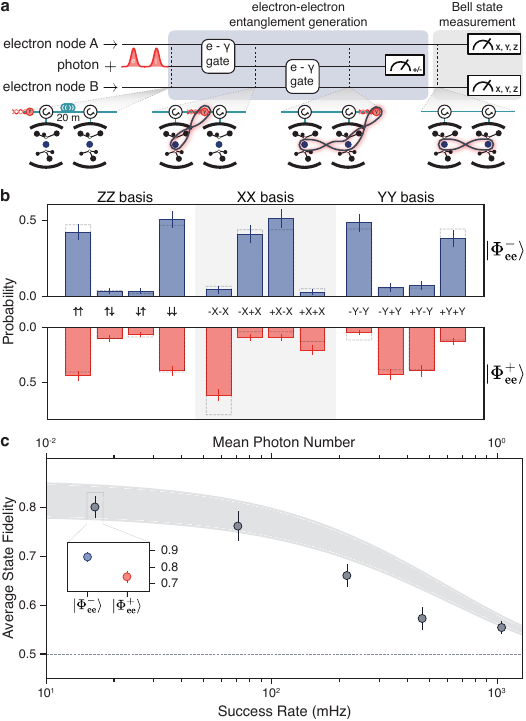}

    \caption{\textbf{Remote entanglement between two electronic spins.} \textbf{a.} Entanglement generation sequence. A photonic qubit is entangled with the electron spin in node A using the $e-\gamma$ gate. A second $e-\gamma$ gate entangles the photonic qubit with node B, generating a GHZ state among the two electronic qubits and the photonic qubit. A measurement of the photonic qubit in the $\ket{\pm}$ basis heralds the generation of an electronic Bell state  $\ket{\Phi_{\text{ee}}^\pm}$. \textbf{b.} Measurement results of Bell state measurement. Measured correlations in the ZZ, XX, and YY bases of the electronic spin corresponding to a Bell state fidelity of   $\mathcal{F}_{\ket{\Phi_{\text{ee}}^-}} = 0.86(3)$  (blue) and $\mathcal{F}_{\ket{\Phi_{\text{ee}}^+}} = 0.74(3)$  (red). Dashed bars show correlations predicted by a theoretical model using independently measured performance parameters of our system. \textbf{c.} Sweep of mean photon number of the photonic qubit showing that the success rates can be increased by sending photonic qubits with a higher mean photon number. The average fidelity of the generated $\ket{\Phi_{\text{ee}}^+}$ and $\ket{\Phi_{\text{ee}}^-}$ states is plotted. Inset shows fidelities of states shown in b. Entanglement is shown to persist above the classical limit (dashed line) for success rates up to $\SI{1}{\hertz}$.  Filled curves show predictions by a theory model using independently measured performance parameters of our system \cite{SI}.   Error-bars in b. and c. are one s.d.}
    \label{fig:2}
\end{figure}

To demonstrate the basic principles of network operation, we first focus on the nodes connected directly via a $\approx \SI{20}{\meter}$ long optical fiber and employ electro-optical frequency shifting (see \citen{Methods} for more details). The above protocol is applied using weak coherent states (WCS, with mean photon number $\mu = 0.017$) to encode time-bin qubits. After the TDI measurement heralds the generation of a Bell state, single qubit rotations and subsequent readout of the electron spin at each node implement the measurement of the correlations $\braket{\sigma^A_{i} \sigma^B_{i}}, i \in \{x,y,z\}$, which we abbreviate as XX, YY, and ZZ, respectively. \cref{fig:2} b shows the results of the correlation measurements, from which we extract the fidelities of the resulting electron-electron state with respect to the maximally entangled Bell states $\mathcal{F}_{\ket{\Phi_{\text{ee}}^-}} = 0.86(3)$ (if the TDI measured $\ket{-}$), and $\mathcal{F}_{\ket{\Phi_{\text{ee}}^+}} = 0.74(3)$ (if the TDI measured $\ket{+}$), unambiguously demonstrating entanglement between the two nodes. The observed difference in fidelity is due to one source of infidelity associated with the imperfect reflection contrast of the two cavity-coupled SiVs. This results in reflection of the photonic qubit even when the electron spin is in the low-reflectivity $\ket{\downarrow_{\text{e}}}$ state. For our system configuration, this type of error accumulates preferentially for the $\ket{\Phi_{\text{ee}}^+}$ state, which is why $\mathcal{F}_{\ket{\Phi_{\text{ee}}^+}}$  is consistently lower than $\mathcal{F}_{\ket{\Phi_{\text{ee}}^-}}$ \cite{SI}. Further error sources include contributions from 2- or higher photon number Fock states of the WCS used as time-bin photonic qubits. By varying the mean photon number $\mu$ in the WCS, we can increase the entanglement generation rate at the cost of reduced fidelity of the generated state. We explore this trade-off in \cref{fig:2} c, where we show that we are able to operate at success rates of \SI{1}{\hertz} while maintaining entanglement.

\subsection{Nuclear spin entanglement}

\begin{figure}[hbt!]
    \centering
    \includegraphics[width=1\linewidth]{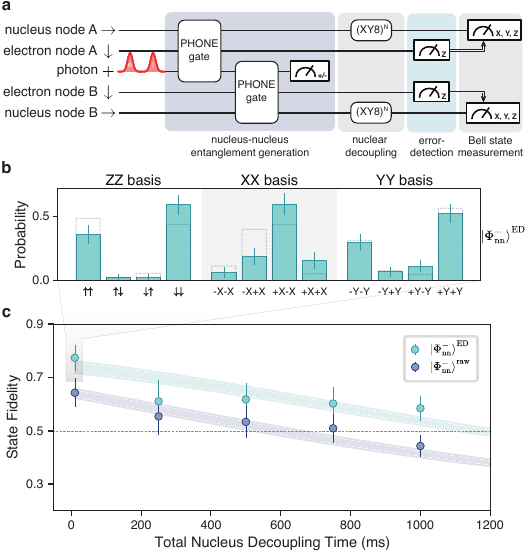}

    \caption{\textbf{Remote entanglement and long-lived storage using nuclear spins.} \textbf{a.} Entanglement generation and subsequent dynamical decoupling using nuclear spin qubits. Nuclear-nuclear entanglement is created by sequentially entangling a time-bin photonic qubit with the $^{29}$Si nuclei at nodes A and B using two PHONE gates. Measurement of the electron spin qubits allows for integrated error detection by flagging MW gate errors occurred during the PHONE gate. 
    \textbf{b.} Results of Bell state measurement of $\ket{\Phi_{\text{nn}}^-}$ after performing error detection, resulting in a Bell state fidelity of $\mathcal{F}^{\text{ED}}_{\ket{\Phi_{\text{nn}}^-}} = 0.77(5)$. Dashed bars show correlations predicted by a theoretical model using independently measured performance parameters of our system. \textbf{c.} Decoherence protection of remotely entangled nuclear-nuclear Bell states, both with (turquoise) and without (blue) error detection. By performing XY8 dynamical decoupling sequences on the two nuclei, entanglement can be preserved for up to \SI{1}{\second}. Filled curves show predictions by a theory model using independently measured performance parameters of our system\cite{SI}. The XY8-1 decoupling sequence was used for the datapoint with \SI{10}{\milli \second} decoupling time while the XY8-128 sequence was used for all other measurements. The dashed line indicates the classical limit. Error-bars in b and c are one s.d.   }
    \label{fig:3}
\end{figure}

Extending remote entanglement to larger distances requires the ability to preserve entanglement long enough such that the heralding signal obtained at node B can be classically relayed to node A. The coherence times of the electron spins in Node A and B  are \SI{125}{\micro \second} and \SI{134}{\micro \second}, respectively. Assuming classical communication via optical fibers in the telecommunication band, the electron spin's decoherence would limit the distance between the nodes to approximately $\SI{25}{\kilo \meter}$. To overcome this limitation, we demonstrate remote entanglement generation between two $^{29}$Si nuclei, which are long-lived quantum memories with storage times of over \SI{2}{\second}\cite{doi:10.1126/science.add9771}.
Analogously to the generation of electron-electron entanglement, remote nuclear entanglement is mediated by the photonic time-bin qubit (\cref{fig:3} a). Thus, the first step of the remote entanglement generation sequence is creating entanglement between a photonic time-bin qubit and the $^{29}$Si nuclear spin at node A. This is achieved using the recently demonstrated PHOton-Nucleus Entangling (PHONE) gate, which only uses microwave pulses to directly entangle the $^{29}$Si nuclear spin with the photonic qubit \cite{Methods, doi:10.1126/science.add9771}, without the need to swap quantum information from electron to nuclear spin. After applying the PHONE gate on the SiV in node A and the photonic qubit, in the ideal limit, their quantum state is:

\begin{equation*}
\ket{\text{Photon}, \text{SiV A}} = \left(\ket{e \downarrow^{A}_{\text{n}}} + \ket{l \uparrow^{A}_{\text{n}}} \right) \ket{\downarrow^{A}_{\text{e}}} / \sqrt{2} .
\end{equation*}

This implies that unless a microwave gate error has occurred, the electron spin is disentangled from the nuclear spin and is in the $\ket{\downarrow^{A}_{\text{e}}}$ state. Thus, the electron spin can be used as a flag qubit to perform error detection by discarding a measurement when the electron spin is measured in $\ket{\uparrow^{A}_{\text{e}}}$. By performing a second PHONE gate between the $^{29}$Si nuclear spin of Node B and the time-bin qubit, and by subsequently measuring out the photonic time-bin qubit in the $\ket{\pm}$ basis, the nuclear Bell states $\ket{\Phi_{\text{nn}}^\pm}$ are created. Following the entanglement generation, we perform XY8-type decoupling sequences on both nuclei to protect the nuclear-nuclear Bell state from decoherence caused by a quasi-static environment. \cref{fig:3} b shows the probability correlations of the resulting $\ket{\Phi_{\text{nn}}^-}$ state using a XY8-1 decoupling sequence with a total nuclear spin decoupling time of \SI{10}{\milli \second}. After using error detection by discarding measurements where the electronic flag qubits are measured in the $\ket{\uparrow_{\text{e}}}$ state, the Bell state fidelity is 
$\mathcal{F}^{\text{ED}}_{\ket{\Phi_{\text{nn}}^-}} = 0.77(5)$, which is an improvement from the directly measured value of $\mathcal{F}^{\text{raw}}_{\ket{\Phi_{\text{nn}}^-}} = 0.64(5)$ without error detection. Similar to $\ket{\Phi_{\text{ee}}^+}$, the generated $\ket{\Phi_{\text{nn}}^+}$ state accumulates errors due to imperfect reflectance contrast, see \citen{SI} for more information. \cref{fig:3} c shows Bell state fidelities for longer total nuclear decoupling times. By performing XY8-128 decoupling sequences, entanglement can be preserved for up to \SI{500}{\milli \second}, with the application of error detection further extending this to one second. 

\subsection{Entanglement distribution through 35 km of deployed fiber}

\begin{figure*}[hbt!]
    \centering
    \includegraphics[width=1\linewidth]{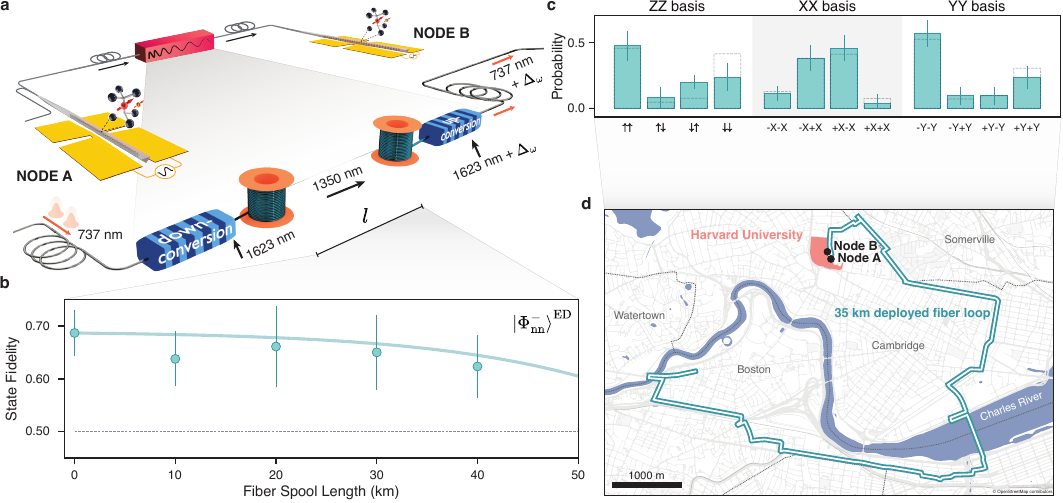}
    \caption{\textbf{Nuclear spin entanglement distribution through 35 km of deployed fiber.} \textbf{a.} Schematic of quantum frequency conversion setup. At node A, the photonic qubit is downconverted from \SI{737}{\nano \meter} to \SI{1350}{\nano \meter}, which can propagate with low loss in telecom single mode fibers. At the node B, it is upconverted back to \SI{737}{\nano \meter}. The pump laser frequencies in the up- and downconversion setups are detuned by $\Delta_{\omega} = \SI{13}{\giga \hertz}$ to compensate for the difference in optical frequencies of the two SiVs. \textbf{b.} Nuclear spin Bell state fidelities for varying lengths of telecom fiber spools between the two nodes. Entanglement persists for fiber lengths up to \SI{40}{\kilo \meter}. Bell state decoherence can be explained by a model incorporating a decrease in signal-to-noise ratio due to dark counts at \SI{2.7}{\hertz} and conversion noise photons at \SI{2.5}{\hertz} (solid line). The dashed line shows the classical limit. \textbf{c.} Measurement results of Bell state measurement of $\ket{\Phi_{\text{nn}}^-}^{\text{ED}}$ state created through a \SI{35}{\kilo \meter} long deployed fiber link shown in d., resulting in a fidelity of $\mathcal{F}^{\text{ED}}_{\ket{\Phi_{\text{nn}}^-}} = 0.69(7)$. Dashed bars show correlations predicted by a theoretical model using independently measured performance parameters of our system. \textbf{d.} Route of the deployed fiber link connecting node A and node B. It consists of \SI{35}{\kilo \meter} deployed telecom fiber routed towards and back from an off-site location,  crossing four municipalities in the greater Boston metropolitan region. Error-bars in b. and c. are one s.d.   }
    \label{fig:4}
\end{figure*}

Light at the SiV's resonant wavelength  (\SI{737}{\nano \meter}) experiences a high in-fiber loss of $>\SI{4}{\decibel \per \kilo \meter}$, which limits the range of remote entanglement distribution at this wavelength. 
In order to make our quantum network compatible with existing classical communication infrastructures that utilize low-loss optical fibers, we employ bidirectional QFC to and from the telecom O-band (\cref{fig:4} a) \cite{bersin2023telecom, SI}. After the photonic qubit at \SI{737}{\nano \meter} is reflected off node A's SiV, a fiber-coupled, periodically poled lithium niobate (PPLN) waveguide pumped with \SI{1623}{\nano \meter} light converts the photonic qubit's wavelength to \SI{1350}{\nano \meter}. This frequency lies in the telecom O-band and shows low attenuation  ($<\SI{0.3}{\decibel \per \kilo \meter}$) in conventional telecom single-mode fiber. After downconversion, the photonic qubit is sent through telecom fiber of varying length before a second PPLN upconverts the photonic qubit back to \SI{737}{\nano \meter}. This bi-directional frequency conversion allows for straightforward bridging of the frequency difference $\Delta_{\omega}$ of the two SiVs: the pump laser's frequency of the upconversion setup is offset by $\Delta_{\omega}$ from the downconversion pump laser's frequency. The total efficiency of the bidirectional QFC, including a final filter cavity, is \SI{5.4}{\percent}, while the noise counts at node B's superconducting nanowire single-photon detector (SNSPD) are \SI{2.5}{\hertz}. \\

Using this frequency conversion scheme together with the entanglement method described above (\cref{fig:3} a), we remotely entangle two $^{29}$Si nuclei through spools of low-loss telecom fiber up to \SI{40}{\kilo \meter} in length (\cref{fig:4} b). For future repeater node applications of truly space-like separated quantum network nodes, it is important that entanglement persists until all nodes have received the classical heralding signal. To account for this effect,  we execute an XY8-1 decoupling sequence for a total duration of \SI{10}{\milli \second} before performing the Bell state measurement. The decoupling duration is much larger than the classical signal traveling time $\Delta t (l) \approx \SI{200}{\micro \second}$ for the maximal fiber length of $l = \SI{40}{\kilo \meter}$. Thus, for the measured fiber distances, Bell state decoherence does not impact the measured Bell state fidelities. Instead, we find that the fiber-distance-dependence of the nuclear-nuclear entanglement fidelities is well described by SNSPD dark counts and telecom conversion noise photons, which reduce the signal-to-noise ratio at high fiber attenuation (solid line in \cref{fig:4} b).\\

In a practical setting, large-scale quantum networks can strongly benefit from existing fiber infrastructure to allow for long-distance entanglement distribution. Deployed fibers are subject to added noise, and excess loss, as well as phase- and polarization drifts \cite{bersin2023development, bersin2023telecom}. We demonstrate that our system is
compatible with conventional fiber infrastructure and is
 resilient to these error sources by generating nuclear entanglement through a \SI{35}{\kilo \meter} loop of telecom fiber deployed in the Boston area urban environment (see \cref{fig:4} d). The overall measured loss in the loop (\SI{17}{\decibel} at \SI{1350}{\nano \meter}) exceeds the nominal fiber attenuation of \SI{11}{\decibel} at this wavelength, indicative of excess loss typical to deployed environments.  Since the input polarization of the up-converting PPLN needs to align with the dipole moment of the crystal,  polarization drifts introduced by the deployed fiber are actively compensated in order to prevent a loss in conversion efficiency \cite{Methods}. Using the deployed link, we generate entanglement with a fidelity of $\mathcal{F}^{\text{ED}}_{\ket{\Phi_{\text{nn}}^-}} = 0.69(7)$ (\cref{fig:4} c), demonstrating the quantum network performance in a realistic fiber environment. 

\subsection{Outlook}

Our experiments demonstrate key ingredients for building large-scale deployed networks using the SiV-based integrated nanophotonic platform. They open the door for exploration of a variety of quantum networking applications, ranging from distributed blind quantum computing\cite{drmota2023verifiable} and non-local sensing, interferometry and clock networks \cite{Khabiboulline_2019PRL, Nichol2022}, to the generation of complex photonic cluster states \cite{Thomas2022}. Extension to entanglement distribution between true space-like separated nodes using deployed fiber requires only relatively minor experimental modifications and is not limited by the performance of the quantum nodes \cite{SI}. The success rate of the entanglement generation is currently limited by losses in the bi-directional QFC,
 which can be minimized by improving mode-matching into the PPLN and the efficiency of the filtering setup \cite{Wang2023}. Furthermore, in-fiber attenuation could be further reduced to \SI{0.2}{\decibel \per \kilo \meter} by using two-stage QFC to \SI{1550}{\nano \meter}\cite{schäfer2023twostage}. The use of WCS also reduces the success rate and fidelity, which could be avoided by using SiV-based single photon sources \cite{Knall_2022} combined with active strain tuning of the nanophotonic cavities for wavelength matching
 \cite{srujanstrain, Machielse2019}. Efficient coupling between the fiber network and the nanophotonic cavity could be improved by recently demonstrated cryogenic packaging techniques  \cite{zeng2023cryogenic}, while cooling requirements of the repeater nodes could be eased by deterministic straining of SiVs \cite{danielstrain}.  Entanglement fidelities could be improved by working with previously demonstrated nanophotonic cavities with higher cooperativity\cite{Bhaskar2020}. Implementing the above improvements, electron-electron entanglement fidelities of $\sim 0.95$ with success rates of $\sim$ \SI{100}{\hertz} could be achieved \cite{SI}. Finally, the number of accessible qubits could be increased by addressing weakly coupled $^{13}$C spins \cite{PhysRevX.9.031045}, allowing for more flexible multi-node network configurations. Combining these advances with the potential ability to create a large number of cavity-QED systems fabricated on a chip, this approach can eventually result in large-scale, deployable quantum networking systems.

\bibliographystyle{custom_bst/naturemag.bst}
\bibliography{refs}

\begin{thebibliography}{10}
\expandafter\ifx\csname url\endcsname\relax
  \def\url#1{\texttt{#1}}\fi
\expandafter\ifx\csname urlprefix\endcsname\relax\def\urlprefix{URL }\fi
\providecommand{\bibinfo}[2]{#2}
\providecommand{\eprint}[2][]{\url{#2}}

\bibitem{Kimble2008}
\bibinfo{author}{Kimble, H.~J.}
\newblock \bibinfo{title}{The quantum internet}.
\newblock \emph{\bibinfo{journal}{Nature}} \textbf{\bibinfo{volume}{453}},
  \bibinfo{pages}{1023--1030} (\bibinfo{year}{2008}).

\bibitem{Briegel1998}
\bibinfo{author}{Briegel, H.~J.}, \bibinfo{author}{D{\"{u}}r, W.},
  \bibinfo{author}{Cirac, J.~I.} \& \bibinfo{author}{Zoller, P.}
\newblock \bibinfo{title}{{Quantum repeaters: The role of imperfect local
  operations in quantum communication}}.
\newblock \emph{\bibinfo{journal}{Physical Review Letters}}
  \textbf{\bibinfo{volume}{81}}, \bibinfo{pages}{5932--5935}
  (\bibinfo{year}{1998}).

\bibitem{Childress_2006}
\bibinfo{author}{Childress, L.}, \bibinfo{author}{Taylor, J.~M.},
  \bibinfo{author}{S{\o}rensen, A.~S.} \& \bibinfo{author}{Lukin, M.~D.}
\newblock \bibinfo{title}{Fault-tolerant quantum communication based on
  solid-state photon emitters}.
\newblock \emph{\bibinfo{journal}{Physical Review Letters}}
  \textbf{\bibinfo{volume}{96}}, \bibinfo{pages}{070504}
  (\bibinfo{year}{2006}).

\bibitem{doi:10.1126/science.aam9288}
\bibinfo{author}{Wehner, S.}, \bibinfo{author}{Elkouss, D.} \&
  \bibinfo{author}{Hanson, R.}
\newblock \bibinfo{title}{Quantum internet: A vision for the road ahead}.
\newblock \emph{\bibinfo{journal}{Science}} \textbf{\bibinfo{volume}{362}},
  \bibinfo{pages}{eaam9288} (\bibinfo{year}{2018}).

\bibitem{azuma2023quantum}
\bibinfo{author}{Azuma, K.} \emph{et~al.}
\newblock \bibinfo{title}{Quantum repeaters: From quantum networks to the
  quantum internet}.
\newblock \emph{\bibinfo{journal}{Rev. Mod. Phys.}}
  \textbf{\bibinfo{volume}{95}}, \bibinfo{pages}{045006}
  (\bibinfo{year}{2023}).

\bibitem{RevModPhys.74.145}
\bibinfo{author}{Gisin, N.}, \bibinfo{author}{Ribordy, G.},
  \bibinfo{author}{Tittel, W.} \& \bibinfo{author}{Zbinden, H.}
\newblock \bibinfo{title}{Quantum cryptography}.
\newblock \emph{\bibinfo{journal}{Rev. Mod. Phys.}}
  \textbf{\bibinfo{volume}{74}}, \bibinfo{pages}{145--195}
  (\bibinfo{year}{2002}).

\bibitem{PhysRevLett.67.661}
\bibinfo{author}{Ekert, A.~K.}
\newblock \bibinfo{title}{Quantum cryptography based on bell's theorem}.
\newblock \emph{\bibinfo{journal}{Physical Review Letters}}
  \textbf{\bibinfo{volume}{67}}, \bibinfo{pages}{661--663}
  (\bibinfo{year}{1991}).

\bibitem{Monroe_2014}
\bibinfo{author}{Monroe, C.} \emph{et~al.}
\newblock \bibinfo{title}{Large-scale modular quantum-computer architecture
  with atomic memory and photonic interconnects}.
\newblock \emph{\bibinfo{journal}{Physical Review A}}
  \textbf{\bibinfo{volume}{89}}, \bibinfo{pages}{022317}
  (\bibinfo{year}{2014}).

\bibitem{ang2022architectures}
\bibinfo{author}{Ang, J.} \emph{et~al.}
\newblock \bibinfo{title}{Architectures for multinode superconducting quantum
  computers.} \bibinfo{pages}{arXiv:2212.06167} (\bibinfo{year}{2022}).

\bibitem{Khabiboulline_2019PRL}
\bibinfo{author}{Khabiboulline, E.~T.}, \bibinfo{author}{Borregaard, J.},
  \bibinfo{author}{Greve, K.~D.} \& \bibinfo{author}{Lukin, M.~D.}
\newblock \bibinfo{title}{Optical interferometry with quantum networks}.
\newblock \emph{\bibinfo{journal}{Physical Review Letters}}
  \textbf{\bibinfo{volume}{123}}, \bibinfo{pages}{070504}
  (\bibinfo{year}{2019}).

\bibitem{Giovannetti_2001}
\bibinfo{author}{Giovannetti, V.}, \bibinfo{author}{Lloyd, S.} \&
  \bibinfo{author}{Maccone, L.}
\newblock \bibinfo{title}{Quantum-enhanced positioning and clock
  synchronization}.
\newblock \emph{\bibinfo{journal}{Nature}} \textbf{\bibinfo{volume}{412}},
  \bibinfo{pages}{417--419} (\bibinfo{year}{2001}).

\bibitem{RevModPhys.87.1379}
\bibinfo{author}{Reiserer, A.} \& \bibinfo{author}{Rempe, G.}
\newblock \bibinfo{title}{Cavity-based quantum networks with single atoms and
  optical photons}.
\newblock \emph{\bibinfo{journal}{Rev. Mod. Phys.}}
  \textbf{\bibinfo{volume}{87}}, \bibinfo{pages}{1379--1418}
  (\bibinfo{year}{2015}).

\bibitem{doi:10.1126/science.aah6875}
\bibinfo{author}{Sipahigil, A.} \emph{et~al.}
\newblock \bibinfo{title}{An integrated diamond nanophotonics platform for
  quantum-optical networks}.
\newblock \emph{\bibinfo{journal}{Science}} \textbf{\bibinfo{volume}{354}},
  \bibinfo{pages}{847--850} (\bibinfo{year}{2016}).

\bibitem{Nguyen2019}
\bibinfo{author}{Nguyen, C.~T.} \emph{et~al.}
\newblock \bibinfo{title}{Quantum network nodes based on diamond qubits with an
  efficient nanophotonic interface}.
\newblock \emph{\bibinfo{journal}{Physical Review Letters}}
  \textbf{\bibinfo{volume}{123}}, \bibinfo{pages}{183602}
  (\bibinfo{year}{2019}).

\bibitem{ourari2023indistinguishable}
\bibinfo{author}{Ourari, S.} \emph{et~al.}
\newblock \bibinfo{title}{Indistinguishable telecom band photons from a single
  erbium ion in the solid state}.
\newblock \emph{\bibinfo{journal}{Nature}} \textbf{\bibinfo{volume}{620}},
  \bibinfo{pages}{977–981} (\bibinfo{year}{2023}).

\bibitem{Ruskuc_2022}
\bibinfo{author}{Ruskuc, A.}, \bibinfo{author}{Wu, C.-J.},
  \bibinfo{author}{Rochman, J.}, \bibinfo{author}{Choi, J.} \&
  \bibinfo{author}{Faraon, A.}
\newblock \bibinfo{title}{Nuclear spin-wave quantum register for a solid-state
  qubit}.
\newblock \emph{\bibinfo{journal}{Nature}} \textbf{\bibinfo{volume}{602}},
  \bibinfo{pages}{408--413} (\bibinfo{year}{2022}).

\bibitem{riedel2023efficient}
\bibinfo{author}{Riedel, D.} \emph{et~al.}
\newblock \bibinfo{title}{Efficient photonic integration of diamond color
  centers and thin-film lithium niobate}.
\newblock \emph{\bibinfo{journal}{ACS Photonics}}  (\bibinfo{year}{2023}).

\bibitem{Molesky2018}
\bibinfo{author}{Molesky, S.} \emph{et~al.}
\newblock \bibinfo{title}{Inverse design in nanophotonics}.
\newblock \emph{\bibinfo{journal}{Nature Photonics}}
  \textbf{\bibinfo{volume}{12}}, \bibinfo{pages}{659--670}
  (\bibinfo{year}{2018}).

\bibitem{parker2023diamond}
\bibinfo{author}{Parker, R.~A.} \emph{et~al.}
\newblock \bibinfo{title}{A diamond nanophotonic interface with an optically
  accessible deterministic electronuclear spin register.}
\newblock \emph{\bibinfo{journal}{Nature Photonics}}  (\bibinfo{year}{2023}).

\bibitem{hofmann2012heralded}
\bibinfo{author}{Hofmann, J.} \emph{et~al.}
\newblock \bibinfo{title}{Heralded entanglement between widely separated
  atoms}.
\newblock \emph{\bibinfo{journal}{Science}} \textbf{\bibinfo{volume}{337}},
  \bibinfo{pages}{72--75} (\bibinfo{year}{2012}).

\bibitem{Daiss_2021}
\bibinfo{author}{Daiss, S.} \emph{et~al.}
\newblock \bibinfo{title}{A quantum-logic gate between distant quantum-network
  modules}.
\newblock \emph{\bibinfo{journal}{Science}} \textbf{\bibinfo{volume}{371}},
  \bibinfo{pages}{614--617} (\bibinfo{year}{2021}).

\bibitem{Ritter2012AnEQ}
\bibinfo{author}{Ritter, S.} \emph{et~al.}
\newblock \bibinfo{title}{An elementary quantum network of single atoms in
  optical cavities}.
\newblock \emph{\bibinfo{journal}{Nature}} \textbf{\bibinfo{volume}{484}},
  \bibinfo{pages}{195--200} (\bibinfo{year}{2012}).

\bibitem{van_Leent_2022}
\bibinfo{author}{van Leent, T.} \emph{et~al.}
\newblock \bibinfo{title}{Entangling single atoms over 33{\hspace{0.167em}}km
  telecom fibre}.
\newblock \emph{\bibinfo{journal}{Nature}} \textbf{\bibinfo{volume}{607}},
  \bibinfo{pages}{69--73} (\bibinfo{year}{2022}).

\bibitem{PhysRevLett.130.050803}
\bibinfo{author}{Krutyanskiy, V.} \emph{et~al.}
\newblock \bibinfo{title}{Entanglement of trapped-ion qubits separated by 230
  meters}.
\newblock \emph{\bibinfo{journal}{Physical Review Letters}}
  \textbf{\bibinfo{volume}{130}}, \bibinfo{pages}{050803}
  (\bibinfo{year}{2023}).

\bibitem{PhysRevLett.124.110501}
\bibinfo{author}{Stephenson, L.~J.} \emph{et~al.}
\newblock \bibinfo{title}{High-rate, high-fidelity entanglement of qubits
  across an elementary quantum network}.
\newblock \emph{\bibinfo{journal}{Physical Review Letters}}
  \textbf{\bibinfo{volume}{124}}, \bibinfo{pages}{110501}
  (\bibinfo{year}{2020}).

\bibitem{PhysRevLett.119.010503}
\bibinfo{author}{Stockill, R.} \emph{et~al.}
\newblock \bibinfo{title}{Phase-tuned entangled state generation between
  distant spin qubits}.
\newblock \emph{\bibinfo{journal}{Physical Review Letters}}
  \textbf{\bibinfo{volume}{119}}, \bibinfo{pages}{010503}
  (\bibinfo{year}{2017}).

\bibitem{Bernien_2013}
\bibinfo{author}{Bernien, H.} \emph{et~al.}
\newblock \bibinfo{title}{Heralded entanglement between solid-state qubits
  separated by three metres}.
\newblock \emph{\bibinfo{journal}{Nature}} \textbf{\bibinfo{volume}{497}},
  \bibinfo{pages}{86--90} (\bibinfo{year}{2013}).

\bibitem{Humphreys_2018}
\bibinfo{author}{Humphreys, P.~C.} \emph{et~al.}
\newblock \bibinfo{title}{Deterministic delivery of remote entanglement on a
  quantum network}.
\newblock \emph{\bibinfo{journal}{Nature}} \textbf{\bibinfo{volume}{558}},
  \bibinfo{pages}{268--273} (\bibinfo{year}{2018}).

\bibitem{PhysRevLett.129.050503}
\bibinfo{author}{Luo, X.-Y.} \emph{et~al.}
\newblock \bibinfo{title}{Postselected entanglement between two atomic
  ensembles separated by 12.5 km}.
\newblock \emph{\bibinfo{journal}{Physical Review Letters}}
  \textbf{\bibinfo{volume}{129}}, \bibinfo{pages}{050503}
  (\bibinfo{year}{2022}).

\bibitem{Yu_2020}
\bibinfo{author}{Yu, Y.} \emph{et~al.}
\newblock \bibinfo{title}{Entanglement of two quantum memories via fibres over
  dozens of kilometres}.
\newblock \emph{\bibinfo{journal}{Nature}} \textbf{\bibinfo{volume}{578}},
  \bibinfo{pages}{240--245} (\bibinfo{year}{2020}).

\bibitem{liu2023multinode}
\bibinfo{author}{Liu, J.-L.} \emph{et~al.}
\newblock \bibinfo{title}{Creation of memory–memory entanglement in a
  metropolitan quantum network}.
\newblock \emph{\bibinfo{journal}{Nature}} \textbf{\bibinfo{volume}{629}},
  \bibinfo{pages}{579--585} (\bibinfo{year}{2024}).

\bibitem{Bhaskar2020}
\bibinfo{author}{Bhaskar, M.~K.} \emph{et~al.}
\newblock \bibinfo{title}{{Experimental demonstration of memory-enhanced
  quantum communication}}.
\newblock \emph{\bibinfo{journal}{Nature}} \textbf{\bibinfo{volume}{580}},
  \bibinfo{pages}{60--64} (\bibinfo{year}{2020}).

\bibitem{doi:10.1126/science.add9771}
\bibinfo{author}{Stas, P.-J.} \emph{et~al.}
\newblock \bibinfo{title}{Robust multi-qubit quantum network node with
  integrated error detection}.
\newblock \emph{\bibinfo{journal}{Science}} \textbf{\bibinfo{volume}{378}},
  \bibinfo{pages}{557--560} (\bibinfo{year}{2022}).

\bibitem{bersin2023telecom}
\bibinfo{author}{Bersin, E.} \emph{et~al.}
\newblock \bibinfo{title}{Telecom networking with a diamond quantum memory}.
\newblock \emph{\bibinfo{journal}{PRX Quantum}} \textbf{\bibinfo{volume}{5}},
  \bibinfo{pages}{010303} (\bibinfo{year}{2024}).

\bibitem{bersin2023development}
\bibinfo{author}{Bersin, E.} \emph{et~al.}
\newblock \bibinfo{title}{Development of a boston-area 50-km fiber quantum
  network testbed}.
\newblock \emph{\bibinfo{journal}{Physical Review Applied}}
  \textbf{\bibinfo{volume}{21}}, \bibinfo{pages}{014024}
  (\bibinfo{year}{2024}).

\bibitem{Hepp2014a}
\bibinfo{author}{Hepp, C.} \emph{et~al.}
\newblock \bibinfo{title}{{Electronic structure of the silicon vacancy color
  center in diamond}}.
\newblock \emph{\bibinfo{journal}{Physical Review Letters}}
  \textbf{\bibinfo{volume}{112}}, \bibinfo{pages}{036405}
  (\bibinfo{year}{2014}).

\bibitem{Pingault2017}
\bibinfo{author}{Pingault, B.} \emph{et~al.}
\newblock \bibinfo{title}{Coherent control of the silicon-vacancy spin in
  diamond}.
\newblock \emph{\bibinfo{journal}{Nature Communications}}
  \textbf{\bibinfo{volume}{8}}, \bibinfo{pages}{15579} (\bibinfo{year}{2017}).

\bibitem{Duan2004}
\bibinfo{author}{Duan, L.~M.} \& \bibinfo{author}{Kimble, H.~J.}
\newblock \bibinfo{title}{{Scalable photonic quantum computation through
  cavity-assisted interactions}}.
\newblock \emph{\bibinfo{journal}{Physical Review Letters}}
  \textbf{\bibinfo{volume}{92}}, \bibinfo{pages}{127902}
  (\bibinfo{year}{2004}).

\bibitem{SI}
 \bibinfo{note}{See Supplementary Information for further clarification and
  discussion.}

\bibitem{Methods}
 \bibinfo{note}{See Methods for further clarification and discussion.}

\bibitem{Barrett_2005}
\bibinfo{author}{Barrett, S.~D.} \& \bibinfo{author}{Kok, P.}
\newblock \bibinfo{title}{Efficient high-fidelity quantum computation using
  matter qubits and linear optics}.
\newblock \emph{\bibinfo{journal}{Physical Review A}}
  \textbf{\bibinfo{volume}{71}}, \bibinfo{pages}{060310(R)}
  (\bibinfo{year}{2005}).

\bibitem{srujanstrain}
\bibinfo{author}{Meesala, S.} \emph{et~al.}
\newblock \bibinfo{title}{Strain engineering of the silicon-vacancy center in
  diamond}.
\newblock \emph{\bibinfo{journal}{Physical Review B}}
  \textbf{\bibinfo{volume}{97}}, \bibinfo{pages}{205444}
  (\bibinfo{year}{2018}).

\bibitem{drmota2023verifiable}
\bibinfo{author}{Drmota, P.} \emph{et~al.}
\newblock \bibinfo{title}{Verifiable blind quantum computing with trapped ions
  and single photons}.
\newblock \emph{\bibinfo{journal}{Phys. Rev. Lett.}}
  \textbf{\bibinfo{volume}{132}}, \bibinfo{pages}{150604}
  (\bibinfo{year}{2024}).

\bibitem{Nichol2022}
\bibinfo{author}{Nichol, B.~C.} \emph{et~al.}
\newblock \bibinfo{title}{An elementary quantum network of entangled optical
  atomic clocks}.
\newblock \emph{\bibinfo{journal}{Nature}} \textbf{\bibinfo{volume}{609}},
  \bibinfo{pages}{689--694} (\bibinfo{year}{2022}).

\bibitem{Thomas2022}
\bibinfo{author}{Thomas, P.}, \bibinfo{author}{Ruscio, L.},
  \bibinfo{author}{Morin, O.} \& \bibinfo{author}{Rempe, G.}
\newblock \bibinfo{title}{Efficient generation of entangled multiphoton graph
  states from a single atom}.
\newblock \emph{\bibinfo{journal}{Nature}} \textbf{\bibinfo{volume}{608}},
  \bibinfo{pages}{677--681} (\bibinfo{year}{2022}).

\bibitem{Wang2023}
\bibinfo{author}{Wang, X.} \emph{et~al.}
\newblock \bibinfo{title}{Quantum frequency conversion and single-photon
  detection with lithium niobate nanophotonic chips}.
\newblock \emph{\bibinfo{journal}{npj Quantum Information}}
  \textbf{\bibinfo{volume}{9}}, \bibinfo{pages}{38} (\bibinfo{year}{2023}).

\bibitem{schäfer2023twostage}
\bibinfo{author}{Schäfer, M.}, \bibinfo{author}{Kambs, B.},
  \bibinfo{author}{Herrmann, D.}, \bibinfo{author}{Bauer, T.} \&
  \bibinfo{author}{Becher, C.}
\newblock \bibinfo{title}{Two-stage, low noise quantum frequency conversion of
  single photons from silicon-vacancy centers in diamond to the telecom
  c-band}.
\newblock \emph{\bibinfo{journal}{Advanced Quantum Technologies}}
  \bibinfo{pages}{2300228} (\bibinfo{year}{2023}).

\bibitem{Knall_2022}
\bibinfo{author}{Knall, E.} \emph{et~al.}
\newblock \bibinfo{title}{Efficient source of shaped single photons based on an
  integrated diamond nanophotonic system}.
\newblock \emph{\bibinfo{journal}{Physical Review Letters}}
  \textbf{\bibinfo{volume}{129}}, \bibinfo{pages}{053603}
  (\bibinfo{year}{2022}).

\bibitem{Machielse2019}
\bibinfo{author}{Machielse, B.} \emph{et~al.}
\newblock \bibinfo{title}{Quantum interference of electromechanically
  stabilized emitters in nanophotonic devices}.
\newblock \emph{\bibinfo{journal}{Physical Review X}}
  \textbf{\bibinfo{volume}{9}}, \bibinfo{pages}{031022} (\bibinfo{year}{2019}).

\bibitem{zeng2023cryogenic}
\bibinfo{author}{Zeng, B.} \emph{et~al.}
\newblock \bibinfo{title}{Cryogenic optical packaging of nanophotonic devices
  with coupling loss {$<1 \text{dB}$}.}
\newblock \emph{\bibinfo{journal}{Applied Physics Letters}}
  \textbf{\bibinfo{volume}{123}}, \bibinfo{pages}{161106}
  (\bibinfo{year}{2023}).

\bibitem{danielstrain}
\bibinfo{author}{Assumpcao, D.~R.} \emph{et~al.}
\newblock \bibinfo{title}{Deterministic creation of strained color centers in
  nanostructures via high-stress thin films.}
\newblock \emph{\bibinfo{journal}{Applied Physics Letters}}
  \textbf{\bibinfo{volume}{123}}, \bibinfo{pages}{244001}
  (\bibinfo{year}{2023}).

\bibitem{PhysRevX.9.031045}
\bibinfo{author}{Bradley, C.~E.} \emph{et~al.}
\newblock \bibinfo{title}{A ten-qubit solid-state spin register with quantum
  memory up to one minute}.
\newblock \emph{\bibinfo{journal}{Physical Review X}}
  \textbf{\bibinfo{volume}{9}}, \bibinfo{pages}{031045} (\bibinfo{year}{2019}).

\bibitem{Nguyen2019a}
\bibinfo{author}{Nguyen, C.~T.} \emph{et~al.}
\newblock \bibinfo{title}{{An integrated nanophotonic quantum register based on
  silicon-vacancy spins in diamond}}.
\newblock \emph{\bibinfo{journal}{Physical Review B}}
  \textbf{\bibinfo{volume}{100}}, \bibinfo{pages}{165428}
  (\bibinfo{year}{2019}).

\bibitem{pylabnet}
\bibinfo{author}{Knaut, C.~M.} \emph{et~al.}
\newblock \bibinfo{title}{{pylabnet - Client-server, python-based laboratory
  software}} (\bibinfo{year}{2021}).

\end{thebibliography}

\crefname{table}{Extended Data Table}{Extended Data Table}
\Crefname{table}{Extended Data Table}{Extended Data Table}

\balancecolsandclearpage

\section{Methods}

\subsection{Spin-photon gates}
The electron spin-dependent cavity reflectance is the building block of the photon-electron entangling gate ($e$ - $\gamma$) gate, and the photon-nucleus entangling (PHONE) gate. The functionality of the two gates can be described in a single node configuration, where a TDI measurement is performed after the node. For the $e$ - $\gamma$ gate (see \cref{fig:spinphoton} a), the electron is initialized in the $\ket{\text{SiV}} = \ket{\rightarrow_{\text{e}}} = (\ket{\downarrow_{\text{e}}} + \ket{\uparrow_{\text{e}}}) / \sqrt{2}$ state, and a photonic time-bin qubit $\ket{+} = (\ket{e} + \ket{l}) / \sqrt{2}$ tuned to the frequency of the maximum reflectance contrast point between $\ket{\downarrow_{\text{e}}}$ and $\ket{\uparrow_{\text{e}}}$ is prepared and sent into the cavity-coupled SiV system. The early time-bin is only reflected off the cavity-QED system if the SiV's electron is in the high-reflectivity state $\ket{\uparrow_{\text{e}}}$, and the state proportional to  $\ket{e \downarrow_{\text{e}}}$ is traced out: 

\begin{equation}
\ket{\text{Photon}, \text{SiV}} = \ket{l} (\ket{\downarrow_{\text{e}}} + \ket{\uparrow_{\text{e}}}) / \sqrt{3}  
+  \ket{e \uparrow_{\text{e}}} / \sqrt{3}.
\end{equation}

A NOT$_{e}$ pulse is applied to the electron between the two time-bins, and the subsequent reflection of the late time bin results in a Bell state between the electron spin and the time-bin photonic qubit:

\begin{equation}
\ket{\text{Photon}, \text{SiV} } =  (\ket{e \downarrow_{\text{e}}} + \ket{l \uparrow_{\text{e}}})/ \sqrt{2}
\end{equation}

The spin-photon entanglement generation is heralded by the successful detection of a photon during the TDI measurement, such that single photons lost in transmission do not degrade the fidelity of the entangled state. The PHONE gate (\cref{fig:spinphoton} b) uses the electron spin to mediate entanglement between the nucleus and the photonic qubit. Before sending the time-bin qubit, the electron-nucleus register is initialized in the $\ket{\downarrow_{\text{e}} \rightarrow_{\text{n}}}$ state. A conditional electron spin flipping MW gate C$_\text{n}$NOT$_{e}$ then prepares the following electron-nucleus entangled state: $
\ket{\text{SiV} } = (\ket{\downarrow_{\text{e}} \downarrow_{\text{n}}} +  \ket{\uparrow_{\text{e}} \uparrow_{\text{n}}} ) / \sqrt{2}$. The remaining sequence is similar to the $e-\gamma$ gate, except with the addition of an unconditional NOT$_{e}$-pulse being applied to the electron between the two time-bins. A final conditional electron spin flipping MW gate  $\overline{\text{C}_\text{n}\text{NOT}_{e}}$  disentangles the electron- and the nuclear spin and results in a Bell state between the nuclear spin and the time-bin photonic qubit:

\begin{equation}
\ket{\text{Photon}, \text{SiV}} = \left(\ket{e \downarrow_{\text{n}}} + \ket{l \uparrow_{\text{n}}} \right) \ket{\downarrow_{\text{e}}} / \sqrt{2}    .
\end{equation}

\subsection{Experimental setup}

The optical path of the experiment consists of multiple stages shown in  \cref{fig:SI_exp_setup_noFC} and \cref{fig:SI_exp_setup_just_FC}. A laser table (\cref{fig:SI_exp_setup_noFC} a) prepares all the required optical fields for control and feedback, as well as for the photonic qubits. The photonic qubit is sent to node A (\cref{fig:SI_exp_setup_noFC} b) where it interacts with the SiV-cavity system. The photonic qubit then continues to the frequency shifting stage (\cref{fig:SI_exp_setup_just_FC} a) or telecom frequency conversion stage (\cref{fig:SI_exp_setup_just_FC} b) depending on the specific experiment, after which it travels to node B (\cref{fig:SI_exp_setup_noFC} c) to interact with the second SiV-cavity system. Finally, the photonic qubit is measured in the $\ket{+}/\ket{-}$ basis with the TDI (\cref{fig:SI_exp_setup_noFC} d). Each SiV at nodes A and B is located in a separate dilution refrigerator (BlueFors BF-LD250) with a base temperature below \SI{200}{\milli \kelvin}. The SiV–cavity system is optically accessed through a tapered fiber.  Lasers at \SI{532}{\nano \meter} are used to stabilize the charge state of the SiVs at each node. The SNSPDs (Photon Spot) are used for SiV state readout, filter cavity locking, and TDI locking. Counts from the SNSPDs are recorded on a time-tagger (Swabian Instruments Time Tagger Ultra). A Zurich Instrument HDAWG8 2.4 GSa/s AWG is used for sequence logic, control of the acousto-optic modulators (AOMs) and phase electro-optic modulators (EOMs), as well as microwave (MW) control. The MW and RF chain is identical for node A and B and is described in \citen{doi:10.1126/science.add9771}.

\subsection{Frequency shifting}
SiVs, like all solid-state qubits, display an inhomogenous distribution in their transition frequencies due to sensitivity to their local environment. In the case of the devices used in this work, we typically see a variation of roughly $\pm 50$ GHz for a SiV formed with a given Si isotope (\cref{fig:SI_inhom}). In particular, the two SiVs used in this work have a difference in their optical transition frequency $\Delta_{\omega} = \omega_B - \omega_A = \SI{13}{\giga \hertz}$,  where $\omega_{A/B}$ is the frequency of the SiV in Node A/B respectively.

In order to bridge this frequency difference and generate entanglement between the SiVs, electro-optic frequency shifting is used (\cref{fig:SI_exp_setup_just_FC} a). Briefly, the initial time-bin photon is generated at a frequency of $\omega_A$. After interacting with the SiV in node A, it is sent through a phase EOM which is driven by a signal generator at a microwave frequency of $\Delta_{\omega}$. As the bandwidth of the optical photon is much narrower than $\Delta_{\omega}$ of the modulator, the frequency spectrum of the output photon can be approximated as a series of optical harmonic frequencies with intensity given by:

\begin{equation}
    I(\omega_A + k\Delta_{\omega}) \propto J_k^2(V_0/V_\pi)
\label{eq:phasemodulator}
\end{equation}

where $k$ is an integer of the frequency harmonic used, $J_k$ is the $k$-th Bessel function of the first kind, $V_\pi$ is the half-wave voltage and $V_0$ is the voltage applied to the modulator. As we are interested in shifting the output photon to a frequency $\omega_B$, which corresponds to the $k=1$ harmonic of the output frequency spectrum, the power of the microwave drive of the EOM is set such that this harmonic is maximized. To filter out unwanted harmonics, the phase EOM's output is sent through a free space Fabry-Pérot cavity (linewidth = \SI{160}{\mega \hertz}, finesse = 312) locked at frequency $\omega_B$. The overall efficiency is approximately $\sim 7.4\%$, given by the product of the maximum EOM sideband power occupancy ($34\%$), the insertion loss of the EOM ($\sim 50\%$), and the insertion loss of the filter cavity including transmission through the cavity and free-space to fiber losses ($\sim 40\%$). We note that this method can be used to shift photons across much larger frequency differences without any additional degradation of efficiency, only limited by the bandwidth of practically available microwave sources and EOMs, and even beyond that using higher harmonics of the output frequency spectrum at the expense of efficiency.

\subsection{Telecom frequency conversion}
To operate over long distances, it is beneficial to transmit photons over lower-loss telecom frequencies. Since SiVs have a natural optical transition at \SI{737}{\nano \meter}, which has an in-fiber loss of $>\SI{4}{\decibel \per \kilo \meter}$, we use quantum frequency conversion to convert it to \SI{1350}{\nano \meter}, which is in the telecom O-band and has a loss of  $<\SI{0.3}{\decibel \per \kilo \meter}$ in SMF28 fiber. 

The conversion was performed using fiber-coupled RPE (reverse photon exchange) LN (Lithium Niobate) waveguide devices (AdvR)  with a pump at \SI{1623}{\nano \meter}. For down-conversion $\SI{737}{\nano \meter} +  \SI{1623}{\nano \meter}  \rightarrow  \SI{1350}{\nano \meter}$, a Difference Frequency Generation (DFG)  device was used with a pump power of 23 dBm at saturation, resulting in overall $33 \%$ conversion efficiency. The up-conversion $\SI{1350}{\nano \meter} +  \SI{1623}{\nano \meter}  \rightarrow  \SI{737}{\nano \meter}$ used a Sum Frequency Generation (SFG) device with a pump power of 25 dBm at saturation, resulting in overall $30 \%$ conversion efficiency. The conversion efficiency in both directions is limited by impedance mismatch between the fiber and waveguide modes. The same pump laser is used for DFG and SFG processes with a frequency offset created by an EOM to compensate for frequency differences $\Delta_{\omega} =\SI{13}{\giga \hertz}$ of the SiVs in two nodes. A home-built Fabry-Pérot cavity (linewidth = \SI{160}{\mega \hertz}, finesse = 312) is used to filter excess noise from the pump at the output of the SFG device. For more details about the telecom setup, see \cref{fig:SI_exp_setup_just_FC} b.   

\subsection{Polarization stabilization}

In-field deployed fibers experience polarization drifts, which need to be compensated due to the polarization sensitivity of the up-converting PPLN used after the deployed fiber link. \cref{fig:polstabi} a shows the setup used to characterize and stabilize the change in the degree of polarization (DOP) introduced by the deployed fiber. At the input of the fiber link, a classical signal from a linearly polarized laser at \SI{1350}{\nano \meter} is sent through the deployed link. At the output of the link, a polarimeter measures the DOP as expressed by the two degrees of freedom ellipticity $\chi \in (-\pi / 4, \pi / 4)$ and azimuth $\psi \in (-\pi / 2, \pi / 2)$. Optical switches are used to connect the deployed fiber link to either the full experiment or the polarization stabilization setup. \cref{fig:polstabi} b shows the histogram of measurements of $\chi$ and $\psi$ over a period of 5 days, without polarization stabilization, showing that the DOP traces out the full phase space over this period. To stabilize the polarization at the output of the fiber link to a well-defined DOP (arbitrarily chosen to be $\chi = \psi = 0$), a fiber-squeezer-based polarization controller is used. During the polarization stabilization sequence, the value of the following cost function is evaluated:

\begin{equation}
C(\chi, \psi) = \sqrt{(\cos(\chi)-1)^2 + (\cos(\psi)-1)^2}
\end{equation}

By employing a gradient-descent algorithm, the optimal control voltages of the polarization controller are determined. The results of this polarization stabilization are shown in \cref{fig:polstabi} c, which shows a histogram of the DOP measured during 5 days of nuclear-nuclear entanglement generation via the deployed fiber. The polarization stabilization routine is interleaved with the entanglement generation experiment and runs during \SI{3}{\percent} of the experimental time, minimally impacting the overall duty cycle. The insertion loss of the polarization stabilization setup is approximately \SI{1.5}{dB}.

\subsection{Success rate}
The success rate $r_{\mathrm{suc}}$ of heralded entanglement at the TDI can be described as:
$ r_{\mathrm{suc}} = \eta R D$, where $\eta$ is the success probability of a heralding event, $R$ is the repetition rate of the experiment and $D$ is the duty cycle of the experiment. The success probability is limited by the photonic link efficiency and protocol-specific loss channels, such as the mean photon number $\mu < 1$ of the photonic qubit's WCS (see \citen{SI} for more information). The repetition rate $R$ is limited by the readout time for the two nodes' electron spin, which is \SI{67}{\micro \second} and \SI{17}{\micro \second} for node A and node B, respectively. The duty cycle $D$ is reduced from its ideal value of $1$ by periodic locking of the TDI and the filter cavity after the frequency shifting or telecom conversion setup, active tracking of the SiV's optical frequency, and, in case of the entanglement generation experiment through the deployed fiber link, polarization stabilization. Furthermore, the application of the green ionization laser to reset the SiV's charge state additionally reduces $D$. By discarding measurements where the spin-dependent reflectance contrast falls below a certain threshold, indicative of large spectral diffusion, $D$ is further reduced.  On average, this contrast threshold for the SiV in node A was set to 1:16, while for the SiV in node B it was set to 1:8. These thresholds resulted in an average  rejection of 23\% of datapoints per dataset. \cref{tab:meth_dutycycle} shows average values for $R, \eta, D, \text{ and } r_{\mathrm{suc}}$ for electron-electron entanglement generation, and nuclear-nuclear entanglement generation. 

\subsection{Acknowledgements}

We thank Denis Sukachev and Eric Bersin for useful discussions and experimental help, Johannes Borregaard, Dirk Englund, Saikat Guha, and P. Benjamin Dixon for useful discussions, and Jim MacArthur for assistance with electronics. This work was supported by the AWS Center for Quantum Networking’s research alliance with the Harvard Quantum Initiative, the National Science Foundation (NSF, Grant No. PHY-2012023), NSF EFRI ACQUIRE (5710004174), CUA (PHY-2317134), AFOSR MURI (FA9550171002 and FA95501610323), and CQN (EEC-1941583). Devices were fabricated at the Harvard Center for Nanoscale Systems, NSF award no. 2025158. Y.Q.H acknowledges support from the A*STAR National Science Scholarship. D.R.A. and E.N.K. acknowledge support from an NSF GRFP No. DGE1745303. M.S. acknowledges funding from the NASA Space Technology Graduate Research Fellowship Program. G. B. acknowledges funding from the MIT Peskoff Graduate Research Fellowship.

\subsection{Author contributions}
  C.M.K., A.S., Y.-C.W., D.R.A., P.J.S., Y.Q.H, M.S., D.L., M.K.B. and M.D.L. planned the experiment. B.M. and E.N.K. fabricated the devices. C.M.K., A.S., Y.-C.W., D.R.A., P.J.S., and Y.Q.H, built the set-up and performed the experiment. C.M.K., A.S., Y.-C.W., D.R.A., P.J.S., Y.Q.H, M.S., and G.B. analyzed the data and interpreted the results. N.S. and M.K.B. coordinated the commissioning of the deployed fiber link. C.D.-E. provided support with the tapered-fiber-optical interface.  All work was supervised by H.P., M.L., and M.D.L. All authors discussed the results and contributed to the manuscript. C.M.K., A.S., Y.-C.W., D.R.A., and P.J.S. contributed equally to this work.

\subsection{Supplementary information}
Supplementary Information is available for this paper.

\subsection{Data availability}
All data related to the current study are available from the corresponding author upon request.

\subsection{Code availability}
All analysis code related to the current study are available from the corresponding author upon request.

\subsection{Competing interests}
The authors declare no competing interests.

\subsection{Corresponding author}
All correspondence should be addressed to M.D.L.

\balancecolsandclearpagesingle

\setcounter{figure}{0}
\renewcommand{\figurename}{Extended Data Fig.}
\renewcommand{\tablename}{Extended Data Table}

\begin{figure*}
    \centering
    \includegraphics[width=0.7\linewidth]{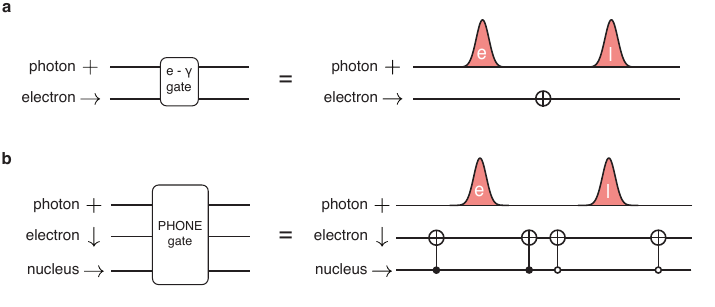}
    \caption{\textbf{Spin-photon gate quantum circuits. a.} Gate representation of the $e$-$\gamma$ gate, entangling a time-bin photonic qubit with the SiV's electron spin. \textbf{b.} Gate representation of the PHONE gate, entangling a time-bin photonic qubit with the SiV's $^{29}$Si nuclear spin.}
    \refstepcounter{EDfig}
    \label{fig:spinphoton}
\end{figure*}

\begin{figure*}
    \centering
    \includegraphics[width=\linewidth, trim=0cm 10cm 0cm 0cm]{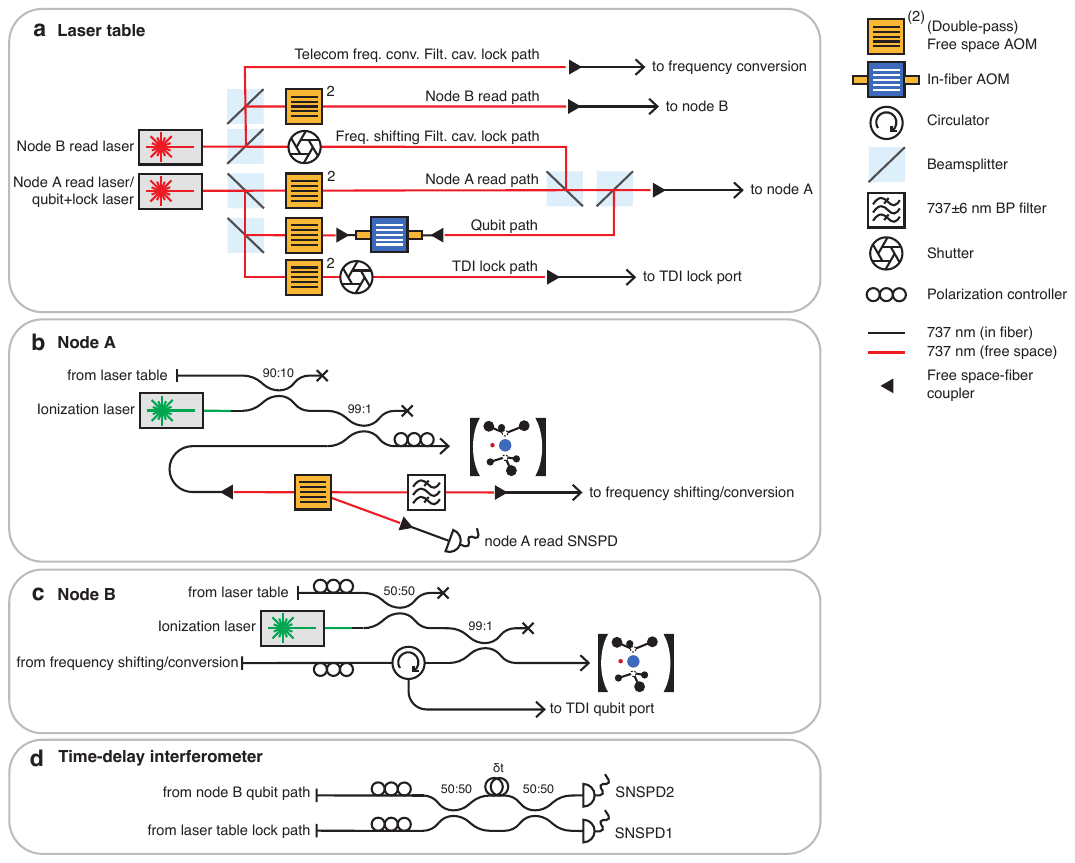}
    \caption{\textbf{Schematic drawing of the two-node quantum network. a.} Laser table. Here, all the optical fields are prepared for node A and B readout, filter cavity, and time-delay interferometer (TDI) locking, and the photonic qubits are shaped by an in-fiber AOM. \textbf{b. and c.} Node A and B. The nodes contain the photonic qubit travel path and a readout path for individual spin readout, as well as an insertion port for a green laser stabilizing the SiVs' charge state. \textbf{d.} Time-delay interferometer. The TDI allows measurement of the photonic qubits in superposition bases.}
    \refstepcounter{EDfig}
    \label{fig:SI_exp_setup_noFC}
\end{figure*}

\begin{figure*}
    \centering
    \includegraphics[width=\linewidth, trim=0cm 15cm 0cm 0cm]{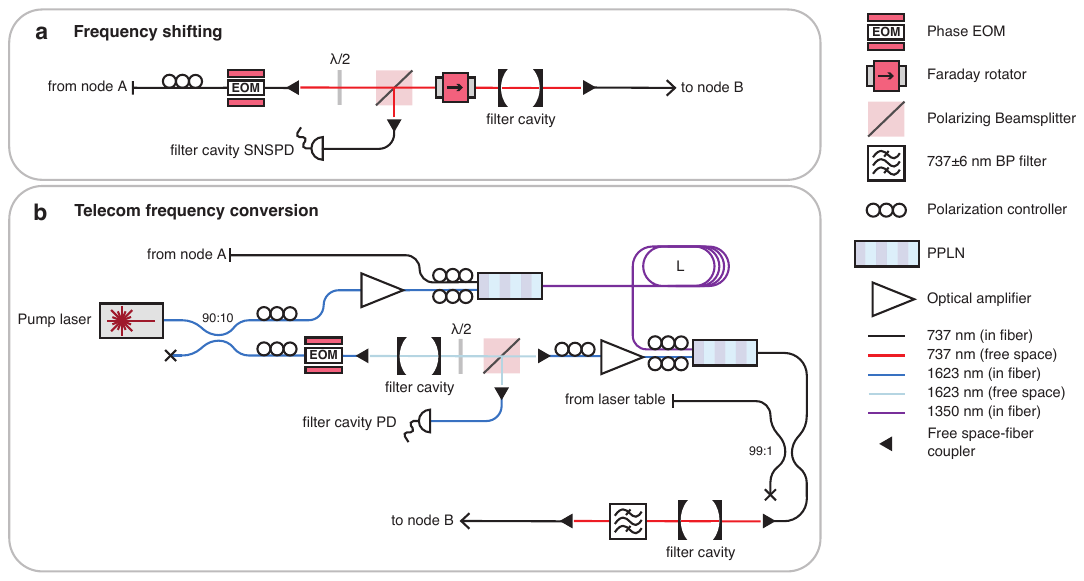}
    \caption{ \textbf{Frequency shifting/conversion setup. a.} Frequency shifting. This setup is used to directly bridge the frequency difference between the two nodes while staying in the visible light range. \textbf{b.} Telecom frequency conversion. This setup is used for long-distance entanglement generation. It contains two frequency conversion stages: one from visible to telecom, and one from telecom to visible. The two stages are separated by a variable amount of fiber spools of total length $L$, or a field-deployed \SI{35}{\kilo \meter} fiber.}
    \refstepcounter{EDfig}
    \label{fig:SI_exp_setup_just_FC}
\end{figure*}

\begin{figure*}
    \centering
    \includegraphics[width=0.8\linewidth, trim=0cm 0cm 0cm 0cm]{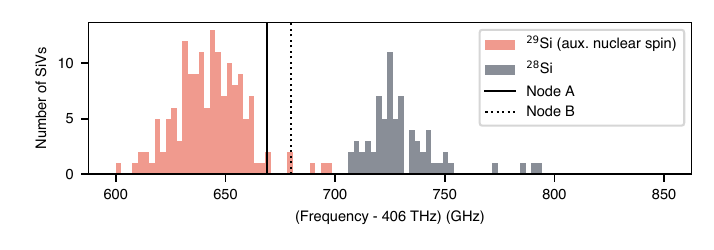}
    \caption{\textbf{Optical resonance frequencies of 223 characterized cavity-coupled SiVs.} The results of this work were produced with SiVs with the $^{29}$Si isotope of silicon, which have a deterministic auxiliary nuclear spin memory. The optical resonance frequencies of the specific SiVs used in this work are given by the solid (node A) and dashed (node B) vertical lines.}
    \refstepcounter{EDfig}
    \label{fig:SI_inhom}
\end{figure*}

\begin{figure*}
    \centering
    \includegraphics[width=0.7\linewidth]{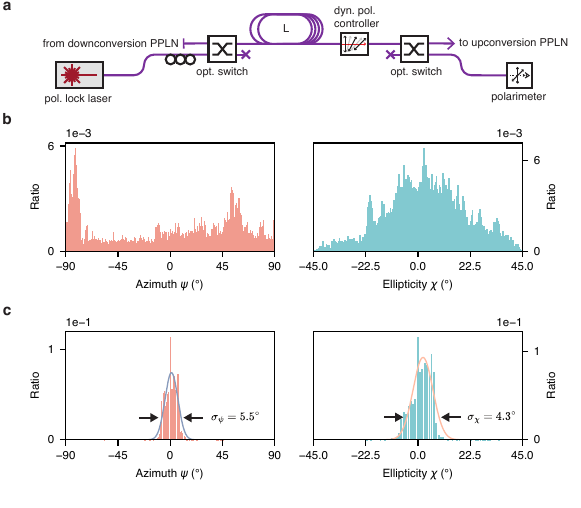}
    \caption{\textbf{Polarization stabilization. a.} Experimental setup used to measure and stabilize the change of the DOP introduced by the \SI{35}{\kilo \meter} fiber link. A linearly polarized lock laser can be launched into the fiber link. After the fiber link, the DOP is measured using a polarimeter. A fiber-based polarization controller is used to stabilize the DOP. \textbf{b.} DOP measurement during 5 hours through the \SI{35}{\kilo \meter} fiber link, not using polarization stabilization. During the measurement period, the initially linearly polarized light traces out the full phase space of the DOP. \textbf{c.} DOP during 5 days of nuclear-nuclear entanglement generation, with polarization stabilization enabled. The DOP can be stabilized at around  $\chi = \psi = 0$.}
    \refstepcounter{EDfig}
    \label{fig:polstabi}
\end{figure*}

{\renewcommand{\arraystretch}{2}

\begin{table*}
\centering
\begin{tabular}{l@{\hskip 0.3in} c c} 

          & Electron-Electron  & Nucleus-Nucleus \\
          & Entanglement & Entanglement (ED)\\

    \hline
    Repetition Rate $R$ & \SI{10}{\kilo \hertz} & \SI{1.4}{\kilo \hertz} \\
    
     Success Probability   $\eta$       &\SI{7.7e-6}{} - \SI{2.5e-4}{} & \SI{2.0e-5}{} \\
     Duty Cycle $D$ & \SI{34}{\percent} & \SI{20}{\percent} \\

     \hline
     Success Rate $r_{\mathrm{suc}}$ & \SI{16}{\milli \hertz} -  \SI{1050}{\milli \hertz} & \SI{6}{\milli \hertz}  \\
    \end{tabular}
     \caption{\textbf{Success rates.} Average values for repetition rate, success probability, and duty-cycle for electron-electron entanglement generation, shown in  \cref{fig:2}, and nucleus-nucleus entanglement generation, shown in \cref{fig:3}. For the nucleus-nucleus entanglement generation, values have been averaged across all decoupling times and $\ket{\Phi_{\text{nn}}^-}^{\text{ED}}$ and $\ket{\Phi_{\text{nn}}^+}^{\text{ED}}$. For the electron-electron entanglement generation, the success probabilities vary due to varying mean photon numbers in the photonic qubit. }
    \label{tab:meth_dutycycle}
\end{table*}}

\clearpage
\onecolumngrid


\setcounter{equation}{0}
\setcounter{figure}{0}
\setcounter{table}{0}
\setcounter{page}{1}
\setcounter{section}{0}
\makeatletter

\renewcommand{\figurename}{Fig.}
\renewcommand{\tablename}{Table}

\renewcommand{\theequation}{S\arabic{equation}}
\renewcommand{\thetable}{S\arabic{table}}
\renewcommand{\thefigure}{S\arabic{figure}}

\crefname{equation}{equation}{equations}
\Crefname{equation}{Equation}{Equations}

\begin{center}
\large \textbf{Supplementary Information for ``Entanglement of Nanophotonic Quantum Memory Nodes in a Telecom  Network"}\\
\normalsize (Dated: May 15, 2024)
\end{center}

\section{I. \hspace{0.2cm} Control Sequence}

The control sequence for SiV control for nuclear-nuclear entanglement generation is summarized in \cref{fig:flow_chart}. Hardware synchronization and MW- and RF-pulse generation are achieved using two Zurich Instrument HDAWG8 2.4 GSa/s AWG (HDAWG). One of the two HDAWGs acts as the controller device and is labeled control system in \cref{fig:flow_chart} a. It provides the central timing sequence, conditional feedback, and AOM control. The second HDAWG acts as worker device, standing by to be triggered by the controller HDAWG to play requested pulses. For more detailed sub-sequences, including electron and nuclear state measurement and initialization, see \cref{fig:flow_chart} b - d.

\section{II. \hspace{0.2cm} cavity-QED parameters}

One of the strengths of the SiV platform is the efficient optical interface. The efficiency of the interface is due to three characteristics of the system. The first one is the strong coupling between the emitter and sub-wavelength mode of the optical cavity, which results in high cooperativity $C = \frac{4g^2}{\kappa_{\text{tot}}\gamma} > 1$. Here,  $g$ is the single-photon Rabi frequency, $\kappa_{\text{tot}}$ the total cavity decay rate, and $\gamma$ is the bare SiV linewidth. Decay rates and linewidths are expressed as full widths at half max (FWHM). The second one is our ability to design and fabricate overcoupled one-sided cavities. And, finally, the third one is the efficient mode matching between the cavity mode and the tapered input fiber.  The reflection amplitude of light for the cavity-coupled SiV at frequency $\omega$ can be expressed as:

\begin{equation}
        \text{Reflection}(\omega)  = 1 - \frac{\kappa_{\text{in}}}{i (\omega - \omega_c) + \kappa_{\text{tot}}/2 + g^2 /(i(\omega-\omega_{SiV})) + \gamma/2} 
\label{eq:cavityQED}
\end{equation}

\begin{figure}
    \centering
    \includegraphics[width=0.75\linewidth, trim=0cm 15cm 0cm 0cm]{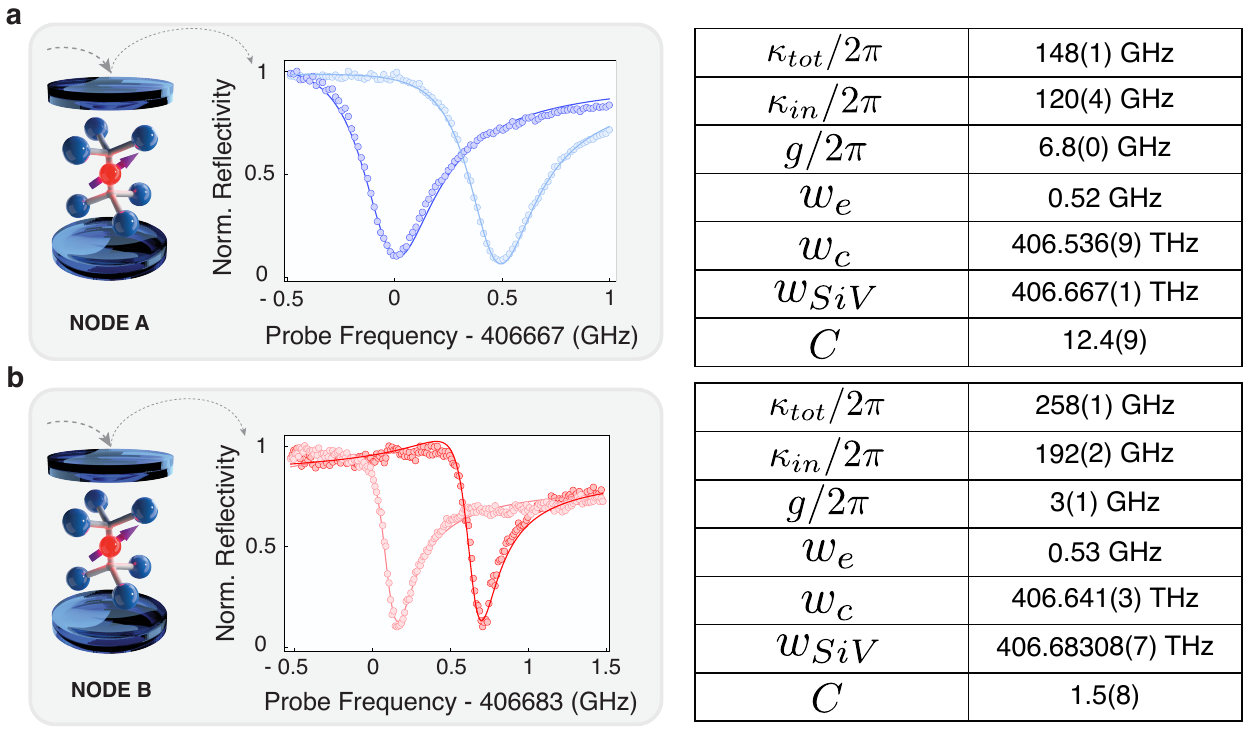}
    \caption{\textbf{Cavity-QED parameters a.} Node A, and \textbf{b.} Node B. Left: Reflection spectra of the cavity-QED system for two electronic spin states. Right: Values of cavity-QED parameters. $\omega_e$ is the frequency splitting between the two electronic spin-states.}
    \refstepcounter{SIfig}
    \label{fig:SI_CavityQED}
\end{figure}

Here,  $\kappa_{\text{in}}$ is the coupling rate of the cavity into the in-coupling port, and $\omega_{SiV (c)}$ is the resonance frequency of the SiV (cavity). The cavity-QED parameters and reflection data for the SiVs used in node A and node B are shown in \cref{fig:SI_CavityQED}. The design and fabrication of the nanophotonic cavity is described in \citen{Knall_2022, Nguyen2019a}.

\section{III. \hspace{0.2cm} Contrast Error Distillation}
\label{sec:ce}
Electron-electron and nuclear-nuclear entanglement are based on state-dependent electron reflectivity. Any reflection of the nominally non-reflective state contributes to Bell state infidelities, an error source we denote contrast error. This error source contributes differently to $\ket{\Phi_{\text{ee}}}^+$  and $\ket{\Phi_{\text{ee}}}^-$, so that selecting for the heralding events for a specific Bell state allows for a higher final Bell state fidelity.

We assume that the reflective electron state has unity reflectivity, while the nominally non-reflective electron state has small non-zero reflectivity $r_A$ ($r_B$) in node A (B). The photon and electrons are prepared in $\ket{+}$ and $\ket{\rightarrow\rightarrow}$, respectively. After performing the $e-\gamma$ gate at the first node, the state becomes

\begin{equation}
    \ket{\gamma, e_A} = \frac{1}{\sqrt{2}}(\ket{e \downarrow^{A}_{\text{e}} } + \ket{l \uparrow^{A}_{\text{e}} }   + r_A ( \ket{e \uparrow^{A}_{\text{e}} } + \ket{l \downarrow^{A}_{\text{e}} }  ) ).
\end{equation}
We use $\gamma$ and $e_{A(B)}$ to denote photonic and electron states in node A (B). The photonic qubit then interacts with the second node, resulting in the state
\begin{equation}
\begin{aligned}
    \ket{\gamma, e_A, e_B} =& \frac{1}{\sqrt{2}}(\ket{e \downarrow^{A}_{\text{e}} \ket{\downarrow^{B}_{\text{e}}} } + \ket{l \uparrow^{A}_{\text{e}} \uparrow^{B}_{\text{e}} }   + r_A ( \ket{e \uparrow^{A}_{\text{e}} \downarrow^{B}_{\text{e}} } + \ket{l \downarrow^{A}_{\text{e}} \uparrow^{B}_{\text{e}} }  ) + \\&
    + r_B ( \ket{e \downarrow^{A}_{\text{e}} \uparrow^{B}_{\text{e}} } + \ket{l \uparrow^{A}_{\text{e}} \downarrow^{B}_{\text{e}} }  ) +  r_A r_B ( \ket{e \uparrow^{A}_{\text{e}} \uparrow^{B}_{\text{e}} } + \ket{l \downarrow^{A}_{\text{e}} \downarrow^{B}_{\text{e}} }  )).
\end{aligned}
\end{equation}
Then, measuring the photon in the $\ket{\pm}$ basis heralds a successful entanglement event. We define $\ket{e_A, e_B}_{\pm}$ as the resulting two-node state after measuring the photon in the $\ket{\pm}$ state:
\begin{equation}
    \begin{aligned}
        \ket{e_A, e_B}_{+} \propto & (1 + r_A r_B) \frac{1}{\sqrt{2}}(\ket{\uparrow^{A}_{\text{e} }\uparrow^{B}_{\text{e}}} + \ket{\downarrow^{A}_{\text{e} }\downarrow^{B}_{\text{e}}}) + \\&
(r_A + r_B) \frac{1}{\sqrt{2}}(\ket{\uparrow^{A}_{\text{e} }\downarrow^{B}_{\text{e}}} + \ket{\downarrow^{A}_{\text{e} }\uparrow^{B}_{\text{e}}}),
    \end{aligned}
\end{equation}
and
\begin{equation}
    \begin{aligned}
        \ket{e_A, e_B}_{-} \propto & (1 - r_A r_B) \frac{1}{\sqrt{2}}(\ket{\uparrow^{A}_{\text{e} }\uparrow^{B}_{\text{e}}} - \ket{\downarrow^{A}_{\text{e} }\downarrow^{B}_{\text{e}}}) + \\&
(r_B - r_A) \frac{1}{\sqrt{2}}(\ket{\uparrow^{A}_{\text{e} }\downarrow^{B}_{\text{e}}} + \ket{\downarrow^{A}_{\text{e} }\uparrow^{B}_{\text{e}}}).
    \end{aligned}
\end{equation}
Here, both states are not normalized. Thus, the infidelity due to the contrast error would then be
\begin{equation}
\begin{aligned}
        \epsilon_+ = (r_A + r_B)^2 / ((r_A + r_B)^2 + (1 + r_Ar_B)^2)\sim (r_A + r_B)^2\\
        \epsilon_- = (r_A - r_B)^2 / ((r_A - r_B)^2 + (1 - r_Ar_B)^2) \sim (r_A - r_B)^2 \\
\end{aligned}
\end{equation}
where the approximation is done by only considering the lowest order errors in $r_A$, $r_B$. We write $r_A = |r_A| e^{i \phi_A}$ and $r_B = |r_B| e^{i \phi_B}$,  and based on the cavity-QED parameters of our system get $\phi_A \approx \phi_B \approx \pi$. In this case, $\epsilon_-$ should be less than $\epsilon_+$ and thus contrast errors in $\ket{e_A, e_B}_{-}$ are suppressed. For nuclear-nuclear entanglement generation, this analysis can be carried out analogously and will also result in a suppression of contrast errors for states heralded by the TDI measuring $\ket{-}$.

\section{IV. \hspace{0.2cm}  Bell State Fidelity and Error Calculation}

Both Bell state fidelity and the sample standard deviation of the Bell state fidelity can be obtained from the measured ZZ, XX, and YY correlators. The Bell state fidelity $\mathcal{F}_{\rho}^{\ket{\Phi^\pm}}$ of an arbitrary quantum state $\rho$ with respect to the Bell states $\ket{\Phi^\pm}$  can be expressed as $\mathcal{F}_{\rho}^{\ket{\Phi^\pm}} = \frac{1}{2}P_{zz} + \frac{1}{4}P_{xx} + \frac{1}{4}P_{yy}$ \cite{doi:10.1126/science.add9771}, with

\begin{align*} 
    P_{zz} &\equiv p_{zz}^{00} + p_{zz}^{11} \\
    P_{xx} &\equiv \begin{cases}
			p_{xx}^{01} + p_{xx}^{10} - p_{xx}^{00} - p_{xx}^{11}, & \text{for $\ket{\Phi^-}$}\\
            p_{xx}^{00} + p_{xx}^{11} - p_{xx}^{01} - p_{xx}^{10}, & \text{for $\ket{\Phi^+}$}
		 \end{cases}    \\
    P_{yy} &\equiv \begin{cases}
			p_{yy}^{00} + p_{yy}^{11} - p_{yy}^{01} - p_{yy}^{10}, & \text{for $\ket{\Phi^-}$}\\
            p_{yy}^{01} + p_{yy}^{10} - p_{yy}^{00} - p_{yy}^{11}, & \text{for $\ket{\Phi^+}$.}
		 \end{cases}
\end{align*}

Here, $p_{nn}^{ij}$, $nn \in  \{zz, xx, yy \}$ describe the probabilities obtaining the various measurement outcomes $(i,j) \in \{ 0,1 \}^2$ for each measurement basis. Now, the sample variance $\sigma^2_{\mathcal{F}}$ of the Bell state fidelity can be expressed as $\sigma^2_{\mathcal{F}} = {\left(\frac{1}{2}\sigma_{zz}\right)}^2 +{\left(\frac{1}{4}\sigma_{xx}\right)}^2 + {\left(\frac{1}{4}\sigma_{yy}\right)}^2 $, with $\sigma^2_{nn \in  \{zz, xx, yy \}}$ describing the sample variance of $P_{{nn \in  \{zz, xx, yy \}}}$. Using the fact that $P_{{nn \in  \{zz, xx, yy \}}}$ follows a binominal distribution, we can use the following expression of the sample variance of a binominally distributed variable with success probability $p$: $\sigma^2(N, p) = \frac{p(1-p)}{N}$, where $N$ is the sample size. Noting that for $P_{zz}$, $p = P_{zz}$, while for $P_{xx}$ and $P_{yy}$, $p = \frac{P_{nn \in \{xx, yy \} } + 1}{2}$, we can finally express the sample standard deviation of the Bell state fidelity as:

\begin{equation}
\sigma_{\mathcal{F}} = \sqrt{\frac{1}{4} \frac{P_{zz} (1-P_{zz})}{N_{zz}} + \frac{1}{16}\frac{(1+P_{xx}) (1-P_{xx})}{N_{xx}} + \frac{1}{16}\frac{(1+P_{yy}) (1-P_{yy})}{N_{yy}}}
\end{equation}

Here, $N_{nn \in  \{zz, xx, yy \}}$ is the sample size for the measurements in the zz, xx, and yy basis.

\section{V. \hspace{0.2cm}  Fidelity Budgets}

\Cref{tab:el-el-err} and \Cref{tab:nu-nu-err} summarize the error contributions from various sources. The individual-node errors, optical contrasts, and TDI locking performance can be directly measured, and their contributions to our target entangled states are analyzed by numerical simulation.

\subsection{V.1. \hspace{0.2cm} Electron-electron entanglement}

\begin{figure}
    \centering
    \includegraphics[width=0.7\linewidth, trim=0cm 0cm 0cm 0cm]{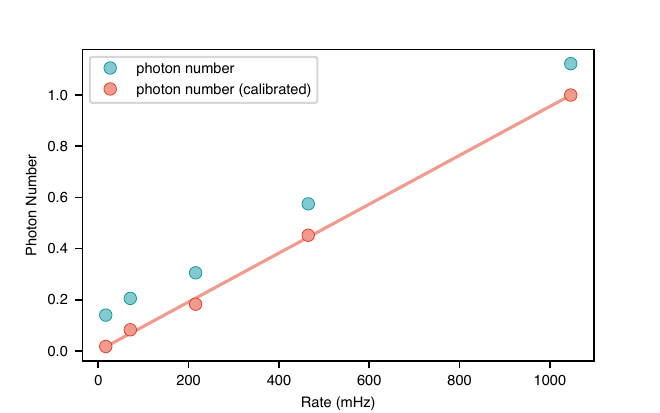}
    
    \caption{\textbf{Average photon number per photonic qubit versus success rate.} The plot shows uncalibrated values (blue), and values where the MW-pulse infidelity has been compensated (red). The mean photon number is measured at node A's SiV. The solid line corresponds to the fitted linear relationship between $\mu$ and the success rate and is used for error simulations.}
    \refstepcounter{SIfig}
    \label{fig:SI_mufitted}
\end{figure}

The numerical simulation and error-budget analysis here are performed in two steps: extracting the average photon numbers per photonic qubit ($\mu$) of experiments and running a simulation model for a given photon number. For photon number extraction, the analysis is carried out by observing the electron spin population resulting from a $X$ basis e-$\gamma$ gate: the electron spin should end up in $\ket{\downarrow_{\text{e}}}$ if no photon arrives, while it ends up being in an equally mixed state of $\ket{\uparrow_{\text{e}}}$ and $\ket{\downarrow_{\text{e}}}$ if $\geq 1$ photon arrives. Based on the coherent state's Poisson distribution, we can then estimate $\mu$. We subtract the infidelity resulting from imperfect MW gates, correcting for the offset observed in the $\mu$-rate plot (\cref{fig:SI_mufitted}). We then feed this photon number into a model that considers four different sources of errors: multiple photon error, imperfect optical contrast, matter-qubit operation error, and TDI locking error. The results are shown in \cref{tab:el-el-err}.

\subsection{V.2. \hspace{0.2cm} Nucleus-nucleus entanglement}

Here, the analysis is carried out similarly as in the above subsection, with two differences: microwave pulse errors and nuclear readout assignment errors. MW pulse errors contribute differently due to the difference between $e$-$\gamma$ and PHONE gate. Furthermore, first-order MW errors are detected with the electron flag qubit, resulting in a lower MW error contribution. To read out nuclear states, a C$_n$NOT$_e$ gate is applied in the middle of two electron readouts, with the measured nuclear state depending on whether the electron state has flipped or not. However, using this approach, the readout fidelity is affected by C$_n$NOT$_e$ gate fidelity, which results in a nuclear readout assignment error. The results of the error analysis are shown in \cref{tab:nu-nu-err}. 

\section{VI. \hspace{0.2cm} Efficiency Breakdown}
The optical losses in our system are broken down in \Cref{tab:SI_eff}. The losses can be seen to come from three major categories. The first is conversion loss to overcome the inhomogeneous distribution, either via visible electro-optic frequency shifting or telecom frequency conversion. The former can be improved through more complex electro-optic shifting schemes, such as serrodyning with high-bandwidth, low-voltage modulators, or shifting using coupled rings. The latter can be improved through further optimization of the conversion setup. Beyond this, the integration of direct strain control of the individual SiVs could be used to tune their transitions and through this overcome the inhomogeneous distribution without any form of photon conversion.

The second major category of losses is our spin photon gate efficiencies, which together lead to almost an order of magnitude penalty on the efficiency. The largest hit to the individual spin-photon gate efficiency is due to the intrinsic 50$\%$ penalty from carving the final spin-photon entangled state from the initial product state. This can be overcome using more complex spin-photon cavity architectures such as phase-based gates using overcoupled cavities or symmetric cavities where photons are collected both on the reflection and transmission ports.

The final category of loss is the sum of the insertion loss of the rest of the components in the system (fiber coupling, circulator, etc.). Beyond improving the insertion loss of each individual component of the system through additional optimization, the net penalty from these components can be alleviated by changing the path architecture from a serial one as used in this work, to a parallel one where a photon is split into two paths, interacts with both nodes and then is re-interfered and measured to generate entanglement, as done in previous experiments\cite{Humphreys_2018, liu2023multinode, PhysRevLett.119.010503, van_Leent_2022}. This parallel architecture ensures that the loss of certain components that are intrinsic to either of the two nodes (such as fiber coupling) only penalizes the overall efficiency a single time as opposed to the double penalty incurred in the serial architecture. On the other side, moving to a parallel architecture using single-photon schemes would increase the experimental complexity due to the need to stabilize two fiber paths, which is especially challenging in deployed fiber environments.

\Cref{tab:SI_eff} does list the estimated photonic link efficiency, corresponding to the insertion loss of our system, and the final success probability, including the protocol-specific loss channels such as losses due to the spin-photon gates. The success probability also accounts for the efficiency hit due to the mean photon number $\mu$ of the WCS used as input in this work. To improve this aspect of the work, high-efficiency deterministic single photon sources or heralded and fed forward single photon sources made from photon-pair sources could be used.

\section{VII. \hspace{0.2cm} System Improvements}
The generated electronic Bell state error sources can be grouped into four categories: TDI locking error, optical contrast error, MW pulse error, and multi-photon error. 
There is no fundamental lower bound to the TDI locking error, and a straightforward improvement of the TDI design, including passive locking and environmental isolation through integration in a vacuum box, could reduce this error source to near-zero. 
The optical contrast error can be significantly suppressed as well. In principle, infinite contrast (and thus zero contrast error) can be  achieved for any overcoupled SiV-cavity system with a cooperativity greater than one. In practice, SiV optical resonance diffusion and stray reflections can lower the contrast, though higher cooperativities mitigate these effects. With reasonable pre-selection and overcoupled SiV-cavity systems with cooperativities higher than 10, optical contrasts of 1:50 are achievable, which would reduce this error source to $\sim 4 \%$ for $\ket{\Phi_{\text{ee}}^+}$ and $<1\%$ for $\ket{\Phi_{\text{ee}}^-}$.

We further note that the SiV at node B was coupled to a nearby $^{13}$C that caused a reduction in MW pulse fidelities. MW pulses are limited by spectral diffusion, spin-state decoherence, as well as MW driving-induced heating. Another constraint on MW pulses is the $^{29}$Si state-dependent splitting of the electron MW transition \cite{doi:10.1126/science.add9771}. This limits how short a MW pulse can be to selectively drive one transition to about 13 ns to apply C$_\text{n}$NOT$_{e}$ gates, or requires very large Rabi frequencies ($\Omega \gg 33$ MHz) to effectively drive both transitions to apply an unconditional NOT$_{e}$ gate. While MW gate errors of $\sim$ 0.1\% have previously been demonstrated \cite{doi:10.1126/science.add9771}, solid state microscopic environmental noise and frequency-dependent noise differs from emitter to emitter, imposing different error limits for different emitters. However, implementing pulse optimization techniques could allow reproduceable and deterministic reduction of MW gate errors below  1\%.

When using WCS as single photons, the error scales as $\sim \mu$, where $\mu$ is the average photon number of the WCS. To reach increasingly higher fidelities requires the use of increasingly lower photon numbers, which would cause a significant reduction in Bell state generation rate. If we replace the WCS with a single photon source, there would not be a need to sacrifice entanglement generation efficiency for fidelity. In this case, the multi-photon error would scale as the single photon source second order correlation at zero delay $ \sim ~g^{(2)}(0)$. A state-of-the-art single photon source at the frequency and bandwidth of the SiV has previously shown to achieve $ g^{(2)}(0) = 1.68\% $ \cite{Knall_2022}. Further improving this fidelity would require improved single photon sources at the frequency and bandwidth of the SiV.
Compiling all these improvements, fidelities of around 0.95 could be achievable in the near future, with even higher fidelities reachable in the longer term with higher cooperativity SiV-cavity systems and improved single photon sources.

Additionally to using a single photon source, the success rates of entanglement generation can further be improved by improving fiber coupling efficiency, with 95\% coupling efficiency shown in previous work \cite{Bhaskar2020}. Higher cooperativity SiV-cavity systems could also enhance the cavity reflectance to 90\% \cite{Bhaskar2020}. This could further boost the success rate by a factor of 7.7, yielding success rates of $\sim$ \SI{8}{\hertz}. Access to strain tuning would remove the need for frequency shifting, further increasing the success rate by a factor of 13.5, resulting in an success rate rate of \SI{100}{\hertz}.

\section{VIII. \hspace{0.2cm} SiV Properties}
\begin{figure}
    \centering
    \includegraphics[width=0.75\linewidth, trim=0cm 0cm 0cm 0cm]{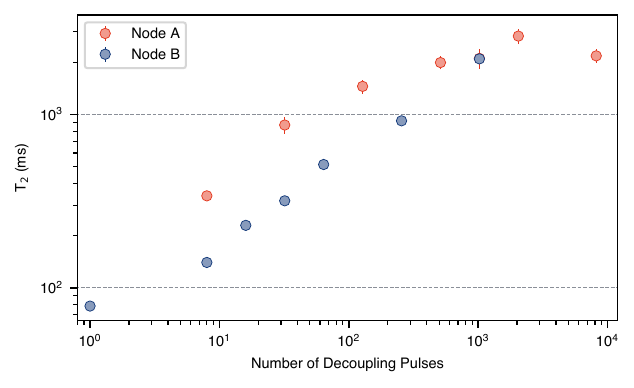}
    \caption{\textbf{T$_2$ coherence times of the $^{29}$Si nuclear spins.} Fitted T$_2$ coherence times of the $^{29}$Si nuclear spin as a function of the number of decoupling pulses for node A and node B. The decoupling sequences are of the form of XY8-N, where N is the number of successive XY8 sequences. Error-bars are one s.d. Dashed lines indicate \SI{0.1}{\second}, and  \SI{1}{\second}.}
    \refstepcounter{SIfig}
    \label{fig:SI_nuc_t2}
\end{figure}

The SiV properties for each node are given in \cref{tab:SI_sivs_params}. The SiVs were picked from a larger measured distribution of SiVs for a combination of desirable device cooperativity, SiV coherence time, and SiV optical stability. The T$_2$ coherence times of the SiV electron and nucleus can be extended with decoupling sequences of increasing length of the form XY8-N, where an XY8 sequence is repeated N times. For a given XY8 sequence, the coherence time of the nuclear spin is about 1000 times higher than for the electron spin due to the smaller nuclear magnetic dipole moment. The maximum achieved T${_2}$ for the nuclear spin is about ~2 seconds for XY8-128 (\cref{fig:SI_nuc_t2}). More XY8 sequences do not seem to further increase the T$_2$, which points to a Markovian noise source limiting the coherence time to about 2 seconds. This is likely due to coupling to phonons through the electron spin.

\section{IX. \hspace{0.2cm} Entanglement Results Additional Data}
\begin{figure}
    \centering
    \includegraphics[width=0.8\linewidth, trim=0cm 0cm 0cm 0cm]{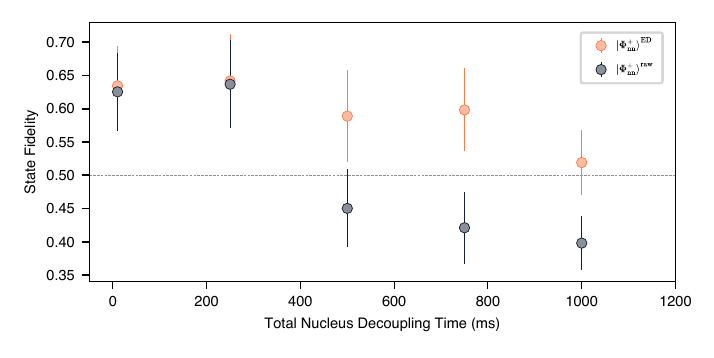}
    \caption{\textbf{$\ket{\Phi_{\text{nn}}^+}$ Bell state fidelities for decoupling experiment}. Fidelities of $\ket{\Phi_{\text{nn}}^+}$ state with (orange) and without (grey) error-detection for different total decoupling times. The dashed line indicates the classical limit. Error-bars are one s.d.}
     \refstepcounter{SIfig}
    \label{fig:SI_plus_hold}
\end{figure}

\begin{figure}
    \centering
    \includegraphics[width=0.8\linewidth, trim=0cm 0cm 0cm 0cm]{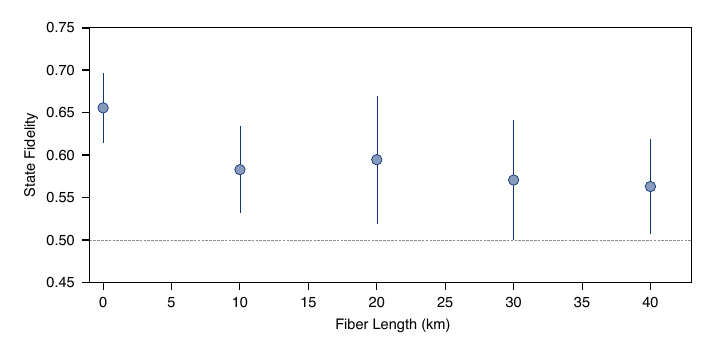}
    \caption{\textbf{$\ket{\Phi_{\text{nn}}^-}$ Bell state fidelities for telecom conversion experiment.} Bell state fidelities of $\ket{\Phi_{\text{nn}}^-}^{\text{raw}}$ state  without error-detection for different fiber spool lengths. The dashed line indicates the classical limit. Error-bars are one s.d.}
    \refstepcounter{SIfig}
    \label{fig:SI_telecom_minus}
\end{figure}

\begin{figure}
    \centering
    \includegraphics[width=0.8\linewidth, trim=0cm 0cm 0cm 0cm]{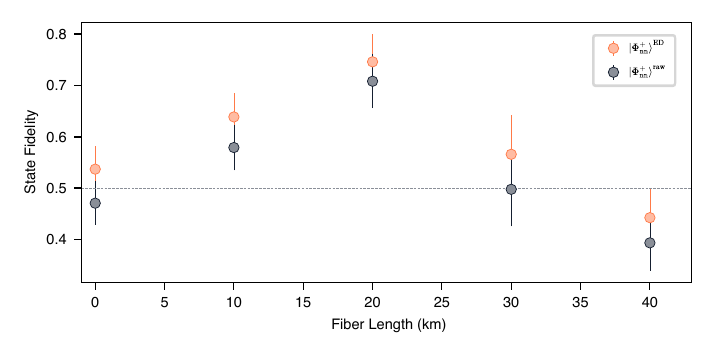}
    \caption{\textbf{$\ket{\Phi_{\text{nn}}^+}$ Bell state fidelities for telecom conversion experiment.} Bell state fidelities of $\ket{\Phi_{\text{nn}}^+}$ state with (orange) and without (grey) error-detection for different fiber spool lengths. The dashed line indicates the classical limit. Error-bars are one s.d.}
     \refstepcounter{SIfig}
    \label{fig:SI_telecom_plus}
\end{figure}

\Cref{fig:SI_plus_hold} shows Bell state fidelities for the $\ket{\Phi_{\text{nn}}^+}$ state for different nuclear decoupling times, both error-detected and raw. The success rates for this measurement are given in 
\cref{table:rates_hold}.  \Cref{fig:SI_telecom_minus} shows the Bell state fidelities of the raw $\ket{\Phi_{\text{nn}}^-}$ state for nuclear-nuclear entanglement generation through spools of up to \SI{40}{\kilo \meter} of low-loss telecom fiber. \Cref{fig:SI_telecom_plus} shows the Bell state fidelities for $\ket{\Phi_{\text{nn}}^+}$ for the same measurement, both error-detected and raw. The success rates for this measurement are given in \cref{table:rates_telecom}. Errors due to imperfect reflectance contrast preferentially affect the $\ket{\Phi_{\text{nn}}^+}$ state, see \hyperref[sec:ce]{section III}. The magnitude of this type of error slowly varies over time due to spectral diffusion of the SiV, macroscopic shifts of the resonance frequency of the nanophotonic cavity, and the laser's frequency stability. These error sources contribute to the dependency of the Bell state fidelities on the fiber length shown in \cref{fig:SI_telecom_plus}.

\section{X. \hspace{0.2cm} Compatibility to True Space-Like Separated Nodes}

\begin{figure}
    \centering
    \includegraphics[width=\linewidth, trim=0cm 0cm 0cm 0cm]{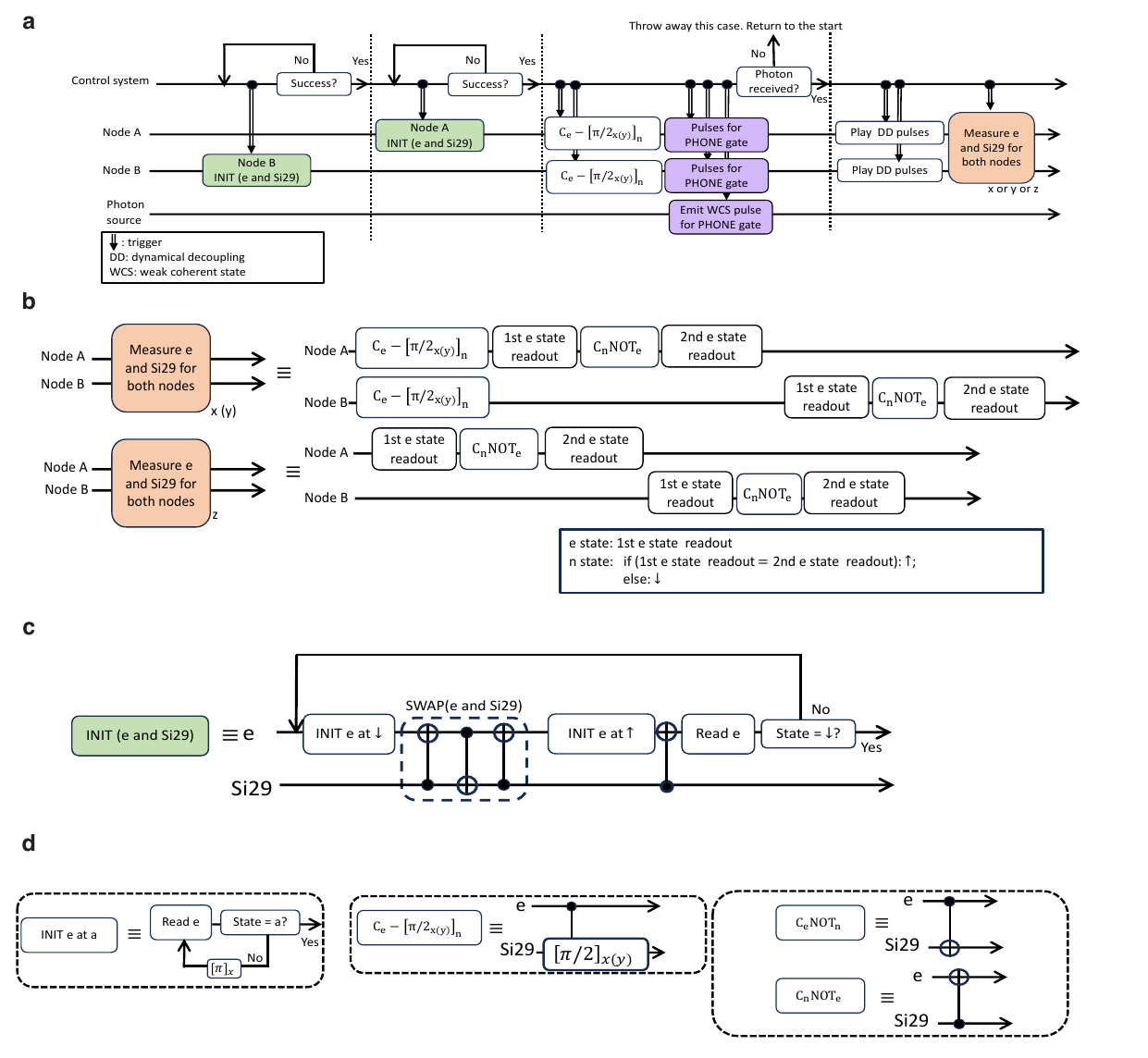}
    \caption{\textbf{Flow chart for experimental control sequence. a.} Summarizes the overall workflow for nucleus-nucleus entanglement through PHONE gates. \textbf{b.} Detailed control sequence for the readout of the electron and the nuclear state. The electron state readout (e state readout in the figure) is performed by detecting photon numbers reflected back from nanophotonic cavities during a read laser pulse. The nuclear state is read by checking if a C$_n$NOT$_e$-pulse flips the electron's state \cite{doi:10.1126/science.add9771}. \textbf{c.} Detailed control sequence of initialization. The nuclear initialization is done by first initializing the electron and then swapping the electron and nuclear state. \textbf{d.} Schematic representations displayed in a.-c. }
    \refstepcounter{SIfig}
    \label{fig:flow_chart}
\end{figure}

Our experimental configuration allows for true space-like separation of the quantum network nodes. The added complexity from the large physical separation can be divided into three categories: synchronization of quantum control hardware, frequency stabilization of lasers, and classical communication of software-timed signals. Our quantum control hardware consists of two AWGs, with one performing the role of the controller AWG, and one as the worker AWG. The worker AWG does not perform any logic operation and is activated using digital signals from the controller AWG. These digital trigger signals could be easily relayed through an additional deployed telecom fiber\cite{bersin2023telecom}. Physically separated nodes also result in the requirement to lock the frequency of two lasers at both nodes. This is particularly important for the laser generating the photonic time-bin qubits at node A, which should match the linewidth and frequency of the laser used to lock the TDI in node B. Such frequency stabilization could be performed by locking both lasers to stable reference cavities located in both nodes. The same strategy could be used for the \SI{1623}{\nano \meter} pump laser used for QFC. Finally, software-timed communication currently is performed using our local area network (LAN) based laboratory control software\cite{pylabnet}. Extending this LAN-based approach to use the internet or a dedicated optical-fiber network is straightforward. It is worth noting that true-space-like separation would decrease the repetition rate of the experiment, as signals indicating successful initialization of node B's nuclear and electronic spin will have to travel classically to node A before the entanglement sequence can begin (see \cref{fig:flow_chart}), and because dynamical decoupling needs to be performed after every entanglement attempt, as opposed to only decoupling after receiving a heralding click, as is done in this work.

\newpage

{\renewcommand{\arraystretch}{1.3}
\begin{table}[h]
\centering
\begin{tabular}{c|c|c|c} 
     Error source & Individual-node error &  Error contribution $\ket{\Phi_{\text{ee}}^-}$& Error contribution $\ket{\Phi_{\text{ee}}^+}$\\ 
    \hline
    Microwave pulse error & $2.5\pm 1.5$ \% (A), $3.5\pm 0.5$ \% (B) & $12.6\pm 5.0$ \% & $7.3 \pm 2.9$ \%\\
    Optical contrast error & $4.3 \pm 1.3$ (A), $8.2 \pm 1.8$ \% (B) & $2.5 
 \pm 0.6$ \% & $13.0 \pm 2.3$ \%\\
    Multi-photon error ($\mu=0.02$) &  - & 0.8 \% & 0.3 \%\\
    TDI locking error &  - & $2.0 \pm 1.0$ \% & $2.0 \pm 1.0$ \%\\
    \hline
    Total expected error &  - & $16.8 \pm 6.4$  \% & $25.9 \pm 5.9$ \%\\
    \hline
    Total observed error &  - & $13.8 \pm 2.9$  \% & $26.0 \pm 3.5$ \%\\
    \hline
    \end{tabular}
    \caption{\textbf{Error budget for electron-electron entanglement.} $\mu$: mean photon number per photonic qubit. Uncertainties for observed errors are one standard deviation. Uncertainties in the remaining rows are estimation ranges.  }
\refstepcounter{SItable}
\label{tab:el-el-err}
\end{table}}

{\renewcommand{\arraystretch}{1.3}
\begin{table}[h]
\centering
\begin{tabular}{c|c|c|c} 
     Error source & Individual-Node Error &  Error contribution $\ket{\Phi_{\text{nn}}^-}^{\text{ED}}$& Error contribution $\ket{\Phi_{\text{nn}}^-}^{\text{raw}}$\\ 
    \hline
    Microwave pulse error & $1.8 \pm 0.3$\% (A), $7.8\pm 0.3$ \% (B) & $\approx 0$ & $22.1 \pm 2.9$ \%\\
    Optical contrast error & $3.5 \pm 0.5$ \% (A), $10.4 \pm 2.1$ \% (B) & $3.9 \pm 0.9$ \% & $3.2 \pm 0.7$ \%\\
    Multi-photon error ($\mu=0.16$) &  - & 6.5 \% & 1.4 \%\\
    TDI locking error &  - & $2.0 \pm 1.0$  \% & $2.0 \pm 1.0$ \%\\
    Nuclear readout assignment error &  $1.5 \pm 0.5$ \% (A), $6.0 \pm1.0$ \% (B) & $4.3 \pm 0.5$ \% & $3.4 \pm 0.4$\%\\
    \hline
    Total expected error &  - & $18.4 \pm 3.1$  \% & $37.0 \pm 5.0$ \%\\
    \hline
    Total observed error &  - & $22.6 \pm 5.0$  \% & $36.7 \pm 5.0$ \%\\
    \hline
    \end{tabular}
    \caption{\textbf{Error budget for nucleus-nucleus  entanglement}.  $\mu$: mean photon number per photonic qubit. Uncertainties for observed errors are one standard deviation. Uncertainties in the remaining rows are estimation ranges.  }
\refstepcounter{SItable}
\label{tab:nu-nu-err}
\end{table}}

{\renewcommand{\arraystretch}{1.3}
\begin{table}[h!]
\centering
\begin{tabular}{l c} 

     & Efficiency \\ 
    \hline
   
    \hline
    Fiber coupling & 60\% (Node A) ${60\%}^2$ (Node B) \\
    Cavity reflectance & 70\% (Node A) 60\% (Node B) \\
    Node A free space setup  & 70 \% \\
    Visible frequency shifting  & 7.4 \% \\
    Telecom frequency conversion & 5.4 \% \\ 
    Circulator & ${70\%}^2$ \\
    SNSPD efficiency & 80\% and 95\% \\
    \hline
    \hline

    {All photonic link efficiency (visible frequency shifting)} & 0.20 \% \\
    {All photonic link  efficiency (telecom frequency conversion)} & 0.15 \% \\

      \hline
    \hline
    Spin-photon gate & 50\%  \\
    $\ket{+}/\ket{-}$ detection & 50\%\\
    WCS mean photon number (variable) & 0.1\\
    \hline
    \hline

    {Success probability $\eta$ (visible frequency shifting)} &  \SI{5.0e-5}{}\\
    {Success probability $\eta$ (telecom frequency conversion)} & \SI{3.7e-5}{} \\

    \end{tabular}
    \caption{\textbf{Estimation of photonic link efficiency and success probability in the two-node quantum network.} Efficiencies with $^2$-superscript enter twice due to our serial network configuration. The photonic link efficiency describes the full insertion loss of our system. The success probability includes the photonic link efficiency and protocol-specific loss channels. Here, the mean photon number of the WCS has been set to a representative value of 0.1.}
    \refstepcounter{SItable}
    \label{tab:SI_eff}
\end{table}}

{\renewcommand{\arraystretch}{1.3}
\begin{table}[h]
\centering
\begin{tabular}{c c c} 

     & Node A & Node B \\ 
    \hline

    $\tau_e$ & 30 ns & 30 ns \\
   
    $\tau_n$ & ${\sim} \SI{15}{\micro\second}$ & ${\sim} \SI{21}{\micro\second}$\\
      
    $T_{2,e}$ XY8-1 & \SI{125}{\micro\second} & \SI{134}{\micro\second} \\
    
    $T_{2,n}$ XY8-1 & 339 ms & 140 ms\\
  
    $T_{2,n}$ XY8-128 & 2.11 s & 2.1 s\\
   
    Magnetic field & 0.39 T &  0.4 T\\
    \hline
    \end{tabular}
    \caption{\textbf{Node A and Node B SiV parameters.} $\tau_{e, (n)}$ is the NOT$_{e, (n)}$-gate duration for the electron (nucleus).}
    \refstepcounter{SItable}
    \label{tab:SI_sivs_params}
\end{table}}

{\renewcommand{\arraystretch}{1.3}
\begin{table}
	\begin{center}
		\begin{tabular}{c|c|c|c}
			Decoupling duration (ms) & Average Rate ED &  Rate $\ket{\Phi_{\text{nn}}^-}^{\text{ED}}$  &  Rate $\ket{\Phi_{\text{nn}}^+}^{\text{ED}}$   \\
			\hline
			10 & 8.3 &  4.3 &  4.0 \\
			250 & 3.4 &  1.6 &  1.8  \\
			500 & 10.0 &  4.8 &  5.2  \\
			750 & 2.0 &  1.0 &  1.0  \\
			1000 & 6.4 &  3.2 & 3.2  \\
		\end{tabular}
	\end{center}
	\caption{\textbf{Rates for decoupling experiment.} Summary of success rates for nuclear-nuclear entanglement generation. Rates are given in mHz.}
                \refstepcounter{SItable}
	\label{table:rates_hold}
\end{table}}

{\renewcommand{\arraystretch}{1.3}
\begin{table}
	\begin{center}
		\begin{tabular}{c|c|c|c}
			Fiber length (km) & Average Rate ED &  Rate $\ket{\Phi_{\text{nn}}^-}^{\text{ED}}$   & Rate $\ket{\Phi_{\text{nn}}^+}^{\text{ED}}$    \\
			\hline
			0 & 8.2 &  3.7 &  4.5  \\
			10 & 5.8 &  2.5 &  3.3  \\
			20 & 2.9 &  1.1 &  1.8  \\
			30 & 5.0 & 2.5 &  2.5  \\
			40 & 1.5 &  0.7 &  0.8 \\
                35 (deployed) & 0.23  & 0.10 &  0.14  \\
		\end{tabular}
	\end{center}
	\caption{\textbf{Rates for telecom conversion experiment.} Summary of success rates for nuclear-nuclear entanglement generation via spools of low-loss telecom fiber and the deployed \SI{35}{\kilo \meter} fiber loop. Rates are given in mHz.}
 \refstepcounter{SItable}
	\label{table:rates_telecom}
\end{table}}

\end{document}